\nonstopmode
\documentclass[preprint,authoryear,12pt]{elsarticle}
\usepackage[T1]{fontenc}
\usepackage{lmodern}
\usepackage[top=1cm,bottom=2cm,left=2cm,right=2cm]{geometry}
\usepackage[utf8]{inputenc}
\usepackage{array}
\usepackage{textcomp}
\usepackage{mathrsfs}
\usepackage{amssymb}
\usepackage{amsmath}
\usepackage{graphicx}
\usepackage[colorlinks=true]{hyperref}

\DeclareMathOperator*{\argmin}{argmin}
\newcommand{\N}{\mathbf{N}}

\newcommand{\R}{\mathbf{R}}

\newcommand{\ud}{\,\text{d}}
\newcommand{\E}{\mathbf{E}}
\newcommand{\Var}{\mathrm{Var }}
\newcommand{\xx}{\mathbf{X}}
\newcommand{\yy}{\mathbf{y}}
\newcommand{\XX}{\mathbf{X}}
\renewcommand{\epsilon}{\varepsilon}

\everymath{\displaystyle\everymath{}}
\providecommand{\abs}[1]{\left\lvert#1\right\rvert}
\providecommand{\norm}[1]{\left\lVert#1\right\rVert}

\usepackage{framed}
\newcounter{algonr}
\newenvironment{algo}{\refstepcounter{algonr}\begin{framed}\textbf{Algorithm \thealgonr: }\par}{\end{framed}}
\journal{Computational Statistics \& Data Analysis}
\begin{document}
\begin{frontmatter}
\title{Confidence intervals for sensitivity indices using reduced-basis metamodels}
\author{A.~Janon\corref{cr}}
\ead{alexandre.janon@imag.fr}
\cortext[cr]{Corresponding author}
\author{M.~Nodet}
\ead{maelle.nodet@inria.fr}
\author{C.~Prieur}
\ead{clementine.prieur@imag.fr}
\address{Joseph Fourier University, Laboratoire Jean Kuntzmann, MOISE team, BP 53, 38041 Grenoble Cedex, France}

\begin{abstract} Global sensitivity analysis is often impracticable for
complex and time demanding numerical models, as it requires a large number of
runs. The reduced-basis approach provides a way to replace the original model by
a much faster to run code. In this paper, we are interested in the information loss induced
by the approximation on the estimation of sensitivity indices. We present a method to provide a robust error assessment, hence enabling significant time savings without sacrifice on precision and rigourousness. We illustrate our method with an experiment where computation time is divided by a factor of nearly 6. We also give
directions on tuning some of the parameters used in our estimation algorithms.
\end{abstract}
\end{frontmatter}

\section*{Introduction}
Many mathematical models use a large number of poorly-known parameters as
inputs. When such models are encountered, it is important for the practitioner
to quantify whether this uncertainty on the inputs has a large repercussion on
the model output. This problem can be tackled by turning the uncertain input
parameters into random variables, whose probability distribution reflects the
practitioner's belief about the oddness of the fact that an input parameter
takes some value. In turn, model output, as function of the model inputs, is a
random variable; its probability distribution, uniquely determined by the
inputs' distribution and the model itself, can give detailed and valuable
information about the behavior of the output when input parameters vary:
range of attained values, mean value and dispersion about the mean (throughout
expectation and standard deviation), most probable values (modes), \emph{etc.} 

\emph{Sensitivity analysis} aims to identify the sensitive parameters, that is
the parameters for which a small variation implies a large variation of the
model output. In \emph{stochastic sensitivity analysis}, one makes use of the
output's probability distribution to define (amongst other measures of
sensitivity) \emph{sensitivity indices} (also known as \emph{Sobol indices}).
Sensitivity index of an output with respect to an input variable is the fraction
of the variance of the output which can be ``explained'' by the variation of the
input variable, either alone (one then speaks about \emph{main effect}), or in
conjunction with other input variables (\emph{total effect}). This way, input
variables can be sorted by the order of importance they have on the output. One
can also consider the part of variance caused by the variation of groups of two
or more inputs, although main effects and total effects are generally sufficient
to produce a satisfying sensitivity analysis. The reader is referred to
\cite{helton2006survey,saltelli-sensitivity} for more information about
uncertainty and sensitivity analysis.

Once these indices have been defined, the question of their effective
calculation remains open. For most models, an exact, analytic computation is not
attainable (even expressing an output as an analytic function of the inputs is
infeasible) and one has to use numerical approximations.

A robust, popular way to obtain such approximations is \emph{Monte Carlo}
estimation. This method simulates randomness in  inputs by sampling a large
number of parameters' values (from the selected inputs' distribution). The model
output is then computed for each sampled value of the parameters. This way, one
obtains a sample of outputs, under the conjugate action of the model and the
input distribution. This sample of outputs is fed into a suitable statistical
estimator of the desired sensitivity index to produce a numerical estimate. The
Monte Carlo approach to computation of Sobol indices is described in
\cite{sobol2001global}, together with improvements in \cite{homma1996importance,saltelli2002making}. 

A major drawback of the Monte Carlo estimation is that a large number of outputs
of the model have to be evaluated for the resulting approximation of the
sensitivity index to be accurate enough. In complex models, where a simulation
for one single value of the parameters may take several minutes, the use of
	these methods ``as-is'' is impracticable. In those cases, one generally makes
	use of a \emph{surrogate model} (also known as \emph{reduced model},
	\emph{metamodel} or \emph{response surface}). The surrogate model has to
	approximate well the original model (called the \emph{full} model), while
	being many times faster to evaluate. The sensitivity index is then calculated
	\emph{via} a sample of outputs generated by a call to the surrogate model,
	thus accelerating the overall computation time. The \emph{reduced-basis} (RB)
	method \cite{nguyen2005certified,grepl2005posteriori,veroy2005certified,grepl2007efficient} is a way of defining surrogate models when
	the original model is a discretization of a partial differential equation
	(PDE) depending on the input parameters. It comes with an \emph{error bound},
	that is, an upper bound on the error between the original output and the
	surrogate output.

The sensitivity index produced by Monte Carlo estimation on a surrogate model
is tainted with a twofold error. Firstly, our Monte-Carlo sampling procedure
assimilates the whole (generally infinite) population of possible inputs with
the finite, randomly chosen, sample; this produces  \emph{sampling}, or
\emph{Monte-Carlo error}. Secondly, using a surrogate model biases the
estimation of the Sobol index, as what is actually estimated is sensitivity
of surrogate output, and \emph{not} the full one; we call this bias the
\emph{metamodel error}.

In order to make a rigorous sensitivity analysis, it is important to assess the
magnitude of these two combined errors on the estimated sensitivity indices.
Such assessment can also be used to help in the choice of correct
approximation parameters (Monte-Carlo sample size and metamodel fidelity) to
achieve a desired precision in estimated indices.

Sampling error can be classically estimated for a moderate cost by using
\emph{bootstrap resampling} \cite{efron1993introduction, archer1997sensitivity}.
Based on statistical estimation theory, the bootstrap technique involves the
generation of a sample of replications of sensitivity index estimator, whose
empirical distribution serves as approximation of the true (unknown) estimator
distribution, in order to produce asymptotic confidence intervals which give
good results in many practical cases. 

A variation on the bootstrap, which addresses sampling error as well as
metamodel error, has been proposed in \cite{storlie2009implementation}; also
\cite{marrel2009calculations} develops a methodology in Kriging metamodels.  In
this paper, we present another confidence interval-based approach for assessing
sampling errors, together with errors caused by reduced-basis metamodels, which
makes use of the certified, \emph{a posteriori} error bound that comes with the
reduced-basis method. \cite{boyaval2009reduced} also makes use of the reduced-basis
output error bound to certify computation of the expectation and the variance of a
model output with neglected sampling error.

The advantages of our approach are: its rigorousness (the impact of the
use of a surrogate model is provably bounded), its efficiency (our bounds are
rather sharp, and go to zero when metamodel errors decrease), its clear
separation between estimation (sampling) and metamodel error, and moderate
computational requirements (time should rather be spent at making a precise
computation than at measuring precision). In other words, our method allows to estimate
sensitivity indices by using 
a reduced basis metamodel which largely speeds up computation times, while rigorously
keeping track of the precision of the estimation.

This paper is organized as follows: in the first part, we go through the
prerequisites for our approach: we give the definition and standard Monte Carlo
estimator of the sensitivity indices we are interested in, and give an overview
of the reduced basis method; in the second and third parts, we present our
confidence interval estimation technique for the sensitivity index, which
accounts for the two sources of error described earlier. In the fourth part, we
present the numerical results we obtain on an
example of a reduced-basis metamodel. 

\section{Model output and sensitivity analysis methodology}
\subsection{Sensitivity indices}
\subsubsection{General setting}
In order to define sensitivity indices, we choose a probability distribution for
the input variables, turning each input variable $X_i$ ($i=1,\ldots,p$) into a
random variable with known distribution; the model output $Y=f(X_1,\ldots,X_p)$
(assumed to be a scalar; multiple outputs can be treated separately) is thus
for $\xx=(X_1,\ldots,X_p)$ a $\sigma(\xx)$-measurable random variable. We further
assume that the $X_i$'s are mutually independent. We also fix throughout all
this paper an input variable of interest $1 \leq i \leq p$. We define the
\emph{first-order main effect} of $X_i$ on $Y$ by:
\begin{equation}
\label{e:si}
S_i = \frac{\Var \E(Y|X_i)}{\Var Y} 
\end{equation}
$S_i$ is the sensitivity index in which we are interested in this paper but
other indices (total effect, high-order effects) exist and our methodology can
readily be extended to these indices.

\subsubsection{Monte-Carlo estimator}
We are interested in the following Monte-Carlo estimator for $S_i$
\cite{homma1996importance,saltelli2002making}: a sample size $N\in\N$ being given, let
$\left\{\xx^k\right\}_{k=1,\ldots,N}$ and $\left\{\xx'^k\right\}_{k=1,\ldots,N}$
be two random i.i.d. samples of size $N$ each, drawn from the distribution of the
input vector $\xx$. 

For $k=1,\ldots,N$, we note:
\[ y_k = f(\xx^k) \]
and:
\[ y_k' = f(X'^k_1,\ldots,X'^k_{i-1},X^k_i,X'^k_{i+1},\ldots,X'^k_p) \]
The Monte-Carlo estimator of $S_i$ is then given by:
\begin{equation}\label{e:mcesti} \widehat{S_i} = \frac{\frac{1}{N} \sum_{k=1}^N y_k y_k' - \left( \frac{1}{N} \sum_{k=1}^N y_k \right) \left( \frac{1}{N} \sum_{k=1}^N y_k' \right) }{  \frac{1}{N} \sum_{k=1}^N y_k^2 - \left( \frac{1}{N} \sum_{k=1}^N y_k \right)^2 } 
\end{equation}

It can be shown that $\widehat{S_i}$ is a strongly consistent estimator of
$S_i$.

\subsection{Metamodel construction: overview of the reduced basis method}
\label{s:RB}
We briefly present here the reduced basis method for affinely parametrized elliptic
partial differential equations. For more details, see \cite{nguyen2005certified}.
\subsubsection{Offline-online decomposition}
The reduced basis method deals with
variational problems of the form: given an input parameter vector $\xx$
belonging to a parameter set $\mathcal D \subset \R^p$,
\begin{equation}
\label{e:pbvar}
\text{find $u(\xx)\in \mathcal{F}$ so that } \;\;\; a(u,v;\xx)=\psi(v) \;\; \forall v \in \mathcal{F} 
\end{equation}
where $\mathcal{F}$ is an appropriate finite dimensional function space (usually a discretization of a continuous function space such as $H^1$ or $H^1_0$), $\psi$ is a linear form on $\mathcal{F}$, and $a(\cdot,\cdot;\xx)$ is a parameter-dependent bilinear form on $\mathcal{F}$ that can be written under \emph{affine form}:
\[ a(v,w;\xx) = \sum_{q=1}^Q \Theta_q(\xx) a_q(v,w) \;\;\; \forall v,w\in \mathcal{F} \]
where $\Theta_q$ are arbitrary real functions defined on $\mathcal D$, and $a_q$ are parameter-independent bilinear functions on $\mathcal{F}$.

Traditional computation of $u(\xx)$ for a prescribed $\xx$ involves looking for $u(\xx)$ as a linear combination:
\[ u(\xx) = \sum_{i=1}^{\dim \mathcal{F}} u_i(\xx) \phi_i \]
where $\left( \phi_i \right)_{i=1,\ldots,\dim \mathcal{F}}$ is a suitable basis of $\mathcal{F}$, and the unknowns are
$(u_i(\xx))_{i=1,\ldots,\dim \mathcal{F}}$. This way, one obtains the following linear system of $(\dim \mathcal{F})$ equations:
\begin{equation}\label{e:linsys} \sum_{i=1}^{\dim \mathcal{F}} \left(\sum_{q=1}^Q \Theta_q(\xx) a_q(\phi_i, \phi_j) \right) u_i(\xx) = \psi(\phi_j) \;\;\; j=1,\ldots,\dim \mathcal{F} \end{equation}
In many cases, the space $\mathcal{F}$ has a large dimension, so as to represent many functions of the continuous function space with a great precision, and the system \eqref{e:linsys} is very large (although one can generally choose $\mathcal{F}$ and $(\phi_i)$ so as to produce a very sparse system).

When \eqref{e:pbvar} has to be solved for many values of $\xx$ (the so-called
\emph{many query} context), the reduced basis method can be used to speed up the
overall computation, which is split into two parts. In the \emph{offline} phase,
we choose a linearly independent family $\mathcal B =
\{\zeta_1,\ldots,\zeta_n\}$ of $n$ vectors in $\mathcal{F}$, and compute and
store the $Q$ $n$-by-$n$ matrices of each $a_q$ form (\emph{ie.} the matrices
$A_q$ whose $(i,j)$ coefficient is $a_q(\zeta_i,\zeta_j)$) and the $n$-vector
representing $\psi$ (\emph{ie.} the vector whose $i$th coefficient is
$\psi(\zeta_i)$) in the basis $\mathcal B$ (called the \emph{reduced basis}).
The offline phase does not depend on a particular value of $\xx$ and can be done
once. Then, for each value of $\xx$ for which $u(\xx)$ has to be computed, one
can proceed to the \emph{online} phase: using information stored during the
offline phase, the following $n$-by-$n$ linear system is assembled and solved
for $(\widetilde u_i(\xx))_{i=1,\ldots,n}$:
\begin{equation} \label{e:linsys2} \sum_{i=1}^n \left(\sum_{q=1}^Q \Theta_q(\xx) a_q(\zeta_i, \zeta_j) \right) \widetilde u_i(\xx) = \psi(\zeta_j) \;\;\; j=1,\ldots,n \end{equation}
Then $\widetilde u(\xx) \approx u(\xx)$ is recovered by using $\widetilde
u(\xx) = \sum_{i=1}^n \widetilde u_i(\xx) \zeta_i$. In many cases, $\{ u(\xx) ;
\xx \in \mathcal D \}$ lies
in a manifold of dimension much smaller than $\dim \mathcal F$, and it is
possible to choose a linear space of approximation of dimension $n \ll \dim \mathcal{F}$ and
thus, solve \eqref{e:linsys2} much faster than \eqref{e:linsys} while keeping
$\widetilde u$ sufficiently close to $u$.  At the end of this section we will see
a method to automatically choose an ``effective'' reduced basis, that allows
accurate representation of $u(\xx)$ for $\xx\in\mathcal D$.

When the model output $f(\xx)$ is a linear functional $f(u(\xx))$ of $u(\xx)$, the surrogate output can be defined as:
\begin{equation}\label{e:surrout} \widetilde f(\xx) := f(\widetilde u(\xx)) = \sum_{q=1}^Q \widetilde u_i(\xx) f(\zeta_i) \end{equation}
whose parameter-independent reals $f(\zeta_i)$ ($i=1,\ldots,n$) can be calculated and stored during the offline phase, allowing evaluation of $\widetilde f(\xx)$ without explicitly forming $\widetilde u(\xx)$, and leading to a metamodel whose complexity of evaluation depends only on its dimension $n$ (and on $Q$) -- and no more on the dimension of the original model $\dim \mathcal{F}$.

\subsubsection{Error bound}
An interesting feature of the reduced basis approach is that it comes with a provable error bound $\epsilon_u(\xx)$ fully computable with a complexity independent of $\dim \mathcal{F}$ \cite{nguyen2005certified}. This error bound satisfies
\[ \norm{ u(\xx) - \widetilde u(\xx) } \leq \epsilon_u(\xx) \;\;\; \forall \xx \in \mathcal D \]
for a chosen Hilbert space norm $\norm{\cdot}$ on $\mathcal{F}$. To present the error bound, we assume, for simplicity, that the $a_q$'s are symmetric bilinear forms, and that $a(\cdot,\cdot;\xx)$ is \emph{uniformly coercive}, that is, $\alpha(\xx)$ defined by:
\[ \alpha(\xx) = \sup_{v\in \mathcal{F}, \norm{v}=1} a(v,v;\xx) \]
satisfies $\alpha(\xx)>0$ for all $\xx\in\mathcal D$. 

We claim that:
\[ \norm{ u(\xx) - \widetilde u(\xx) } \leq \frac{ \norm{ r(\xx) }_{\mathcal{F}'} }{ \alpha(\xx) } \;\;\; \forall \xx \in \mathcal D \]
where $r(\xx)$ is the residual linear form, defined by:
\[ r(\xx) (v) = \psi(v) - a(\widetilde u(\xx),v;\xx) \]
and $\norm{\cdot}_{\mathcal{F}'}$ is the dual norm on $\mathcal{F}$:
\[ \norm{ \ell }_{\mathcal{F}'} = \sup_{v \in \mathcal{F}, \norm{v}=1} \ell(v) \]
Efficient procedures have been developed for efficient offline-online computation of $\norm{ r(\xx) }_{\mathcal{F}'}$, and a \emph{lower bound} $\widetilde\alpha(\xx)<\alpha(\xx)$, leading to a computable error bound on $u$:
\[ \epsilon_u(\xx) = \frac{ \norm{ r(\xx) }_{\mathcal{F}'} }{ \widetilde\alpha(\xx) } \]
This error bound on $u$ can be used to develop an error bound on the output. For example, if $f(\xx)=f(u(\xx))$ and $\widetilde f(\xx)$ is the surrogate output defined in \eqref{e:surrout}, one can use
\begin{equation}\label{e:errc} \epsilon(\xx) = \norm{f}_{\mathcal{F}'} \epsilon_u(\xx) \end{equation}
which satisfies:
\begin{equation}
\label{e:errboundout}
\abs{ f(\xx) - \widetilde f(\xx) } \leq \epsilon(\xx) \;\;\; \forall \xx \in \mathcal D 
\end{equation}
as error bound on the output.

\subsubsection{POD-based procedure for reduced basis choice}
\label{s:podrb}
We now describe a way of selecting a reduced basis $\{\zeta_1,\ldots,\zeta_n\}$.

We randomly choose a finite subset $\Xi = \{\xx^1,\ldots,\xx^m\} \subset\mathcal D$, and compute $u(\xx)$ for each $\xx\in\Xi$. We put the coordinates of $u(\xx)$ with respect to the basis $\{\phi_1,\ldots,\phi_{\dim \mathcal F}\}$ of $\mathcal F$ as columns of a \emph{snapshot matrix} $S$:
\begin{equation}
\label{e:snapmat}
 S = \left( \begin{array}{cccc} u(\xx^1)_1 & u(\xx^2)_1 & \ldots & u(\xx^m)_1 \\
											 u(\xx^1)_2 & u(\xx^2)_2 & \ldots & u(\xx^m)_2 \\
											 \vdots & \vdots & \ldots & \vdots \\
	 u(\xx^1)_{\dim\mathcal F} & u(\xx^2)_{\dim\mathcal F} & \ldots & u(\xx^m)_{\dim\mathcal F} \end{array} \right) 
\end{equation}
We now proceed with the \emph{Proper Orthogonal Decomposition (POD)} of the $S$ matrix: we compute $\{z_1,\ldots,z_n\}$, where $z_i$ is an eigenvector associated with the $i^\text{th}$ largest eigenvalue of the $m$-by-$m$ symmetric matrix $S^T\Omega S$ (where $\Omega$ is the matrix of the scalar product $<,>$ associated with $\norm{\cdot}$, with respect to the $\{\phi_1,\ldots,\phi_{\dim\mathcal F}\}$ basis), and define the vectors of the reduced basis to be:
\[ \zeta_i = \frac{ S z_i }{\norm{ S z_i }} \]
One can show that the $\{\zeta_1,\ldots,\zeta_n\}$ are solutions of the following optimization program:
\[ 
\text{Minimize } \sum_{\xx\in\Xi} \norm{ u(\xx) - \pi[u(\xx)]  }^2, 
\text{ under the constraints } <\zeta_i, \zeta_j>=\left\{ \begin{array}{l} 1 \text{ if } i=j \\																						 0 \text{ else.} \end{array}\right. \]
where $\pi$ is the orthogonal projector onto $\text{Span}(\zeta_1,\ldots,\zeta_n)$.

Proper orthogonal decomposition (also known as Principal component analysis (PCA), or Singular value decomposition (SVD)) \cite{chatterjee2000introduction}, and variants of POD, are widely used in model reduction without error bounds \cite{bui2007goal,bergmann2008numerical}. We also showed in \cite{janon2011certified} that POD reduced bases are efficient with respect to the obtained error bounds.

\subsection{Estimator on reduced model}
Using the reduced model to perform the sensitivity analysis is straightforward: replace every call to the full model $f$ by a call to the reduced one $\widetilde f$. This gives rise to an estimator $\widehat{\widetilde{S_i}}$ converging, for $N\rightarrow+\infty$, to the true value of the sensitivity index on $\widetilde f$:
\[ \widetilde S_i = \frac{\Var \E(\widetilde Y|X_i)}{\Var \widetilde Y} \]
Sampling error of this estimator can be assessed using bootstrap \cite{efron1993introduction}, see Algorithm
\ref{a:bootmc}.

However, doing so does not take in account the gap between $S_i$ and $\widetilde
S_i$. If the metamodel is ``too far'' from the original model, the
$(1-\alpha)$--confidence interval estimated using it will not contain the true
value of $S_i$ with probability close to $1-\alpha$. On the other hand, a
moderate-fidelity metamodel might be well-suited to give a ``rough'' estimate of sensitivity indices -- in some cases such a rough estimate is sufficient -- but the user would like be informed that the metamodel he uses gives him a limited-precision estimator; such a limited precision would reflect in the increase in the width of the output confidence interval.

\section{Quantification of the two types of error in index estimation}
We now present our method for estimating the two types of error that occur in
 Monte-Carlo sensitivity index estimation on a reduced-basis metamodel. In the
 first part \ref{s:bootstrap}, we review the bootstrap, which we will use for
 the treatment of sampling error. In the second part \ref{ss:mmerror}, we show
 how to use reduced-basis bounds to assess metamodel error.

\subsection{Sampling error : bootstrap confidence intervals}
\label{s:bootstrap}
Sampling error, due to the Monte-Carlo evaluation of the variances in
\eqref{e:si}, can be quantified through an
approximate confidence interval calculated using bootstrap
\cite{archer1997sensitivity}. 

We use the bias-corrected (BC) percentile method presented in 
\cite{efron1981nonparametric,efron1986bootstrap}. The principle of this method can be 
summed up the following way: let
$\widehat\theta(X_1,\ldots,X_n)$ be an estimator for an unknown parameter
$\theta$ in a reference population $\mathcal P$. To get a point estimate of
$\theta$, one takes a random i.i.d. $n$-sample $\{ x_1,\ldots,x_n \}$ from $\mathcal
P$, and computes $\widehat\theta(x_1,\ldots,x_n)$. In (nonparametric) bootstrap
we repeatedly, for $b=1,\ldots,B$, sample $\{ x_1[b],\ldots,x_n[b] \}$ with replacement from the original
sample $\{ x_1,\ldots,x_n \}$ and get a \emph{replication} of $\widehat\theta$ by
computing $\widehat\theta[b]=\widehat\theta(x_1[b],\ldots,x_n[b])$. This way we get a sample
$\mathcal R=\{\widehat\theta[1],\ldots,\widehat\theta[B]\}$ of
replications of $\widehat\theta$.  

Now see how this sample can be used to estimate a confidence interval for
$\theta$. We denote by $\Phi$ the standard normal cdf:
\[ \Phi(z) = \frac{1}{\sqrt{2\pi}} \int_{-\infty}^z \exp\left(- \frac{t^2}{2}
\right) \ud t \]
and by $\Phi^{-1}$ its inverse.

Using $\mathcal R$ and the point estimate $\widehat\theta=\widehat\theta(x_1,\ldots,x_n)$, a ``bias correction
constant'' $z_0$ can be estimated:
\[ \widehat{z_0} = \Phi^{-1}\left( \frac{ \#\{\widehat\theta[b] \in \mathcal R \textrm{~s.t.~}
\widehat\theta[b]\leq\widehat\theta \} }{B} \right) \]

Then, for $\beta\in]0;1[$, we define the
``corrected quantile estimate'' $\widehat q(\beta)$:
\[ \widehat q(\beta)=\Phi(2\widehat{z_0}+z_\beta) \]
where $z_\beta$ satisfies $\Phi(z_\beta)=\beta$.

The central BC bootstrap confidence interval of level $1-\alpha$ is then
estimated
by the interval whose endpoints are the $\widehat q(\alpha/2)$ and
$\widehat q(1-\alpha/2)$ quantiles of $\mathcal R$.

This confidence interval has been justified in \cite{efron1981nonparametric} when there exists an increasing transformation $g$,  $z_0\in\R$ and $\sigma>0$ such that $g(\widehat\theta)\sim\mathcal N(\theta-z_0\sigma,\sigma)$ and $g(\widehat\theta^*)\sim\mathcal N(\widehat\theta-z_0\sigma,\sigma)$, where $\widehat\theta^*$ is the bootstrapped $\widehat\theta$, for fixed sample $\{x_1,\ldots,x_n\}$ and (hence) fixed $\widehat\theta=\widehat\theta(x_1,\ldots,x_n)$. In practice, due to the complex analytic expressions \eqref{e:defsimsiM} of the estimators we are going to bootstrap, it seems hard to prove that such a $g$ exists. However, we give in Section \ref{ss:normapprox} empirical evidence that, for the two estimators defined at \eqref{e:defsimsiM}, $g$ can approximatively be chosen as identity.

The full computation method of a BC bootstrap confidence interval for $S_i$ is given in Algorithm~\ref{a:bootmc}.

The key advantage of bootstrapping our sensitivity estimators is that we do not
require supplementary model evaluations to estimate a confidence interval; hence
the computational overhead for getting a confidence interval (versus pointwise
estimation only) remains quite modest.

\subsection{Metamodel error}
\label{ss:mmerror}
\label{s:boundb}
For a pair of samples $\left( \{\xx^k\}_{k=1,\ldots,N},
\{\xx'^k\}_{k=1,\ldots,N} \right)$, we can use our metamodel output $\widetilde f$
and our metamodel error bound $\epsilon$ to compute, for $k=1,\ldots,N$:
\[ \widetilde y_k=\widetilde f(\xx^k),\;\; \widetilde y_k'=\widetilde
f(X'^k_1,\ldots,X'^k_{i-1},X^k_i,X'^k_{i+1},\ldots,X'^k_p) \]
and:
\[ \epsilon_k=\epsilon(\xx^k),\;\; \epsilon_k'=\epsilon(X'^k_1,\ldots,X'^k_{i-1},X^k_i,X'^k_{i+1},\ldots,X'^k_p ) \]
In this section, we find accurate, explicitly and efficiently computable bounds
$\widehat S_i^m$ and $\widehat S_i^M$, depending only on $\widetilde y_k,
\widetilde y_k', \epsilon_k$ and $\epsilon_k'$ so that:
\begin{equation}
\label{e:boundSI}
\widehat S_i^m \leq \widehat S_i \leq \widehat
S_i^M 
\end{equation}
In other words, we want lower and upper bounds on the full model based sensitivity index estimator
$\widehat S_i$ computable from surrogate model calls.

Let:
\[ R(a;\yy,\mu,\mu')=\sum_{k=1}^N \left( y_k' - (a (y_k-\mu)+\mu') \right)^2 \]
where $\yy=(y_1,\ldots,y_N,y'_1,\ldots,y'_N)$ and $\mu,\mu'\in\R$.

By setting first derivative of $R$ with respect to $a$ to zero, making use
of the convexity of $R(\cdot;\yy,\overline y,\overline y')$ and using:
\[ \widehat{S_i} = \frac{\frac{1}{N} \sum_{k=1}^N y_k y_k' - \left( \frac{1}{N} \sum_{k=1}^N y_k \right) \left( \frac{1}{N} \sum_{k=1}^N y_k' \right) }{  \frac{1}{N} \sum_{k=1}^N y_k^2 - \left( \frac{1}{N} \sum_{k=1}^N y_k \right)^2 } \]
one easily shows that:
\[ \widehat S_i = \argmin_{a\in\R} R(a;\yy,\overline y,\overline y') \]
where: $\overline y=\frac{1}{N} \sum_{k=1}^N y_k$ and $\overline y'=\frac{1}{N} \sum_{k=1}^N y_k'$.

In other words, $\widehat S_i$ is the slope of the linear least squares regression of the $\{ y_k' \}_k$ on the $\{ y_k \}$.

Define:
\begin{equation}
\label{e:defrinf}
R_{inf}(a;\widetilde{\yy},\epsilon,\mu,\mu')=\sum_{k=1}^N \left\{
	\inf_{z\in[\widetilde y_k - \epsilon_k ; \widetilde y_k + \epsilon_k],
	z'\in[\widetilde y_k' - \epsilon_k' ; \widetilde y_k' + \epsilon_k']} \left(
	z'-(a(z-\mu)+\mu') \right)^2 \right\} 
\end{equation}
and:
\begin{equation}
\label{e:defrsup}
R_{sup}(a;\widetilde{\yy},\epsilon,\mu,\mu')=\sum_{k=1}^N \left\{
	\sup_{z\in[\widetilde y_k - \epsilon_k ; \widetilde y_k + \epsilon_k],
	z'\in[\widetilde y_k' - \epsilon_k' ; \widetilde y_k' + \epsilon_k']} \left(
	z'-(a(z-\mu)+\mu') \right)^2 \right\} 
\end{equation}
where $\widetilde{\yy}=(\widetilde y_1,\ldots, \widetilde y_N,\widetilde y_1',\ldots,\widetilde y_N')$, $\epsilon=(\epsilon_1,\ldots,\epsilon_N,\epsilon_1',\ldots,\epsilon_N')$.

It is clear that:
\begin{equation}
\label{e:encadr}
 R_{inf}(a;\widetilde{\yy},\epsilon,\mu,\mu') \leq R(a;\yy,\mu,\mu') \leq R_{sup}(a;\widetilde{\yy},\epsilon,\mu,\mu') \;\;\;
\forall a, \mu, \mu' \in \R
\end{equation}

Note that $R$, $R_{inf}$ and $R_{sup}$ are quadratic polynomials in $a$. We name
$\alpha,\beta,\gamma,\alpha_{inf},\beta_{inf},\gamma_{inf},\alpha_{sup},\beta_{sup}$
and $\gamma_{sup}$ their respective coefficients. In other words, we have:
\[ R(a;\yy,\mu,\mu')=\alpha a^2 + \beta a + \gamma \]
\begin{equation}
\label{e:coefrinf}
R_{inf}(a;\widetilde{\yy},\epsilon,\mu,\mu')=\alpha_{inf} a^2 + \beta_{inf} a + \gamma_{inf} 
\end{equation}
\begin{equation}
\label{e:coefrsup}
R_{sup}(a;\widetilde{\yy},\epsilon,\mu,\mu')=\alpha_{sup} a^2 + \beta_{sup} a + \gamma_{sup} 
\end{equation}
These coefficients depend on $\mu$ and $\mu'$ \footnote{as well on $\yy$ (for $\alpha,\beta,\gamma$) and $\widetilde{\yy}$ and $\epsilon$ (for the other coefficients)}. We do not explicitly write this dependence until the last part of our discussion.

Using \eqref{e:encadr} we see that the quadratic function of $a$:
\[ (\alpha_{inf}-\alpha)a^2+(\beta_{inf}-\beta)a+\gamma_{inf}-\gamma \]
is negative or zero; hence it takes a non-positive value for $a=0$, and has a non-positive discriminant:
\begin{eqnarray}
\label{e:poly1}
\gamma_{inf}-\gamma \leq 0 \\
\label{e:poly2}
\left( \beta_{inf}-\beta \right)^2 \leq 4 (\alpha_{inf}-\alpha)(\gamma_{inf}-\gamma)
\end{eqnarray}
As $\left( \beta_{inf}-\beta \right)^2  \geq 0$, Equations \eqref{e:poly1} and \eqref{e:poly2} above imply that $\alpha_{inf}-\alpha \leq 0$, and that:
\[ \beta_{inf}-\delta_{inf} \leq \beta \leq \beta_{inf}+\delta_{inf} \]
for $\delta_{inf}=2 \sqrt{( \alpha_{inf}-\alpha )( \gamma_{inf}-\gamma )}$.

We now suppose that $\alpha_{inf}>0$. As $\alpha_{inf}$ is computable from $\widetilde y_k,
\widetilde y_k', \epsilon_k$ and $\epsilon_k'$, one can practically check
if this condition is met. If it is not the case, our bound can not be used.
We expect that if the metamodel error is not too large, we have
$\alpha_{inf} \approx \alpha$ and, as $\alpha>0$, the hypothesis
$\alpha_{inf}>0$ is realistic.

So, under this supplementary assumption, we have:
\[ \argmin_a R(a;\yy,\mu,\mu')=-\frac{\beta}{2\alpha} \geq - \frac{\beta_{inf}+\delta_{inf}}{2 \alpha_{inf}} \]

Now using the second part of \eqref{e:encadr} and the same reasoning on the non-positive quadratic function of $a$: $R(a;\yy,\mu,\mu')-R_{sup}(a;\widetilde{\yy},\epsilon,\mu,\mu')$, we get that: $\alpha\leq\alpha_{sup}$, and: $\beta_{sup}-\delta_{sup} \leq \beta \leq \beta_{sup}+\delta_{sup}$. Hence,
\[ \argmin_a R(a;\yy,\mu,\mu') \leq - \frac{\beta_{sup}-\delta_{sup}}{2\alpha_{sup}} \]
where $\delta_{sup}=2 \sqrt{(\alpha-\alpha_{sup})(\gamma-\gamma_{sup})}$. This comes without supplementary assumption, because $\alpha_{sup}\geq\alpha$ and $\alpha>0$, as the minimum of $R(\cdot;\yy,\mu,\mu')$ exists.

As we clearly have $\delta_{inf}$ and $\delta_{sup}$ less than (or equal to) $\widehat\delta:=2\sqrt{(\alpha_{inf}-\alpha_{sup})(\gamma_{inf}-\gamma_{sup})}$, we deduce that:
\[ - \frac{\beta_{inf}(\mu,\mu') + \widehat\delta(\mu,\mu')}{2 \alpha_{inf}(\mu,\mu')} \leq \argmin_a R(a;\yy,\mu,\mu') \leq
- \frac{\beta_{sup}(\mu,\mu') - \widehat\delta(\mu,\mu')}{2 \alpha_{sup}(\mu,\mu')} \]
where we have explicited the dependencies in $\mu$ and $\mu'$.

To finish, it is easy to see that we have:
\begin{equation}
\label{e:defp}
\overline{\mathcal P} := [ \overline {\widetilde y} - \overline \epsilon ;
\overline{\widetilde y} + \overline \epsilon ] \ni \overline y
\end{equation}
and:
\begin{equation}
\label{e:defppr}
\overline{\mathcal P}' := [ \overline{\widetilde y'} - \overline \epsilon' ;
\overline{\widetilde y'} + \overline \epsilon' ] \ni \overline y' 
\end{equation}
(where $\overline{\widetilde y}, \overline{\widetilde y'}, \overline{\epsilon}$ and $\overline{\epsilon'}$ denote, respectively, the means of 
$\left( \widetilde y_k \right)_k, \left( \widetilde y_k' \right)_k, \left( \epsilon_k \right)_k$ and $\left( \epsilon_k' \right)_k$) so that:
\[ \min_{\mu \in \overline{\mathcal P}, \mu' \in \overline{\mathcal P}'} \left( - \frac{\beta_{inf}(\mu,\mu') + \widehat\delta(\mu,\mu')}{2 \alpha_{inf}(\mu,\mu')} \right)
 \leq \widehat S_i = \argmin_a R(a;\yy,\mu,\mu') \leq
\max_{\mu \in \overline{\mathcal P}, \mu' \in \overline{\mathcal P}'} \left(- \frac{\beta_{sup}(\mu,\mu') - \widehat\delta(\mu,\mu')}{2 \alpha_{sup}(\mu,\mu')} \right) \]
Hence, \eqref{e:boundSI} is verified with:
\begin{equation}
\label{e:defsimsiM}
 \widehat S_i^m=\min_{\mu \in \overline{\mathcal P}, \mu' \in
\overline{\mathcal P}'} \left( - \frac{\beta_{inf}(\mu,\mu') +
\widehat\delta(\mu,\mu')}{2 \alpha_{inf}(\mu,\mu')} \right),
\;\;\;\;
\widehat S_i^M=\max_{\mu \in \overline{\mathcal P}, \mu' \in \overline{\mathcal
P}'} \left(- \frac{\beta_{sup}(\mu,\mu') - \widehat\delta(\mu,\mu')}{2
\alpha_{sup}(\mu,\mu')} \right) 
\end{equation}
It is clear that $\widehat S_i^m$ and $\widehat S_i^M$ are computable without
knowing the $y_k$s and $y_k'$s.

In practice, we compute approximate values of $\widehat S_i^m$ and
$\widehat S_i^M$ by replacing the $\min$ and $\max$ over $\overline{\mathcal
P}\times\overline{\mathcal P'}$ by the $\min$ and $\max$ over a finite sample
$\Xi \subset \overline{\mathcal P}\times\overline{\mathcal P'}$. See Algorithm
\ref{a:mcerr} for a summary of the entire computation procedure.

\section{Combined confidence intervals and parameters choice}
\subsection{Combined confidence intervals}
\label{s:bootCI}
In the last section, we have seen how to separately assess sampling error and
metamodel error. 
To take both error into account simultaneously, we propose using bootstrap confidence intervals (see Section \ref{s:bootstrap}) by calculating $B$ bootstrap replications of $\widehat S_i^m $ and $\widehat S_i^M $, where, for $b=1,\ldots,B$ each bootstrap pair $\left( \widehat S_i^m [b] ; \widehat S_i^M [b] \right)$ is computed using $ \left( \widetilde y_k \right)_{k\in L_b},
\left( \widetilde y_k' \right)_{k\in L_b}$ as surrogate output samples, and
associated error bounds $\left( \widetilde \epsilon_k \right)_{k\in L_b}, \left(
\widetilde \epsilon_k' \right)_{k\in L_b}$, where $L_b$ is a list of $N$ integers sampled with replacement from $\{1,\ldots,N\}$.

The BC bootstrap confidence interval procedure (see \ref{s:bootstrap}) can then
be used to produce a $1-\alpha$-level confidence interval $[\widehat
S_{i,\alpha/2}^m;\widehat S_{i,1-\alpha/2}^m]$ for $S_i^m$, and a confidence
interval $[\widehat
S_{i,\alpha/2}^M;\widehat S_{i,1-\alpha/2}^M]$ for $S_i^M$. We then take as
combined confidence interval of level $1-\alpha$ for $S_i$ the range $[ \widehat
S_{i,\alpha/2}^m ; \widehat S_{i,1-\alpha/2}^M]$. This interval accounts for
sampling and metamodels error simultaneously.

\subsubsection*{$\epsilon$ sampling}
Optionally, we can introduce a postulated uncertainty on the error bounds
through what we call \emph{$\epsilon$ sampling}. In $\epsilon$ sampling, the
$b^\text{th}$ bootstrap replicates for $\widehat{S_i^m}$ and $\widehat{S_i^M}$
are computed using $\left( \epsilon_k^* \right)_{k\in L_b}, \left(
	{\epsilon'}_k^* \right)_{k\in L_b}$ as error bounds, where
$\epsilon_k^*$ and $ {\epsilon'}_k^*$ are sampled independently from a uniform
distribution in $\left[\eta_k \epsilon_k ; \epsilon_k\right]$ and $\left[\eta_k'
\epsilon_k' ; \epsilon_k'\right]$ , where $\eta_k, \eta_k' \in [0;1]$ are
alleged \emph{effectivities} of our error bound, that is, an indicator of the
ratio between the true errors $\abs{ \widetilde y_k - y_k }$ and $\epsilon_k$
(and between $\abs{ \widetilde y_k' - y_k' }$ and $\epsilon_k'$). Setting
effectiveness close to zero narrows confidence intervals, putting more trust in
the reduced model than in the error bound, which is considered too
pessimistic; on the contrary, effectiveness close to one means that error bound
does not overestimate true error too much and that the error can not be
considered too smaller than it.

The procedure for obtaining confidence intervals is summed up in Algorithm \ref{a:confInt}.

\subsection{Choice of reduced basis size and Monte-Carlo sample size}
\label{s:towards}
When doing a Monte Carlo estimation of sensitivity indices using a reduced basis
metamodel, by means of confidence intervals computed with the strategy described
above, one has to choose two important parameters : the sample size ($N$) and
the number of elements in the reduced basis ($n$). Increasing $N$ and/or $n$
will increase the overall time for computation (because of a larger number of
surrogate simulations to perform if $N$ is increased, or, if $n$ is increased,
each surrogate simulation taking more time to complete due to a larger linear
system to solve). However, increase in these parameters will also improve the
precision of the calculation (thanks to reduction in sampling error for
increased $N$, or reduction in metamodel error for increased $n$). In practice,
one wants to estimate sensitivity indices with a given precision (\emph{ie.} to
produce $(1-\alpha)$-level confidence intervals with prescribed length), and has
no \emph{a priori} indication on how to choose $N$ and $n$ to do so. Moreover,
for one given precision, there may be multiple choices of suitable couples
	$(N,n)$, balancing between sampling and metamodel error. We wish to choose
	the best, that is, the one who gives the smallest computation time. 

The aim of this section is to describe a simple computational model that helps
us in making a good choice of sample size and reduced basis size to produce a
confidence interval of a desired precision.

\subsubsection*{Formulation as a constrained optimization problem}

On the one hand, we evaluate computation time: an analysis of the reduced basis method shows that the most costly operation made during an online evaluation (see Section \ref{s:RB}) is the resolution of a linear system of $n$ equations; this resolution can be done (e.g., by using Gauss' algorithm) with $O(n^3)$ operations. This has to be multiplied by the required number of online evaluations, i.e. the sample size $N$. Hence, we may assume that computation time is proportional to $N \times n^3$.

On the other hand, the mean length of the
$(1-\alpha)$-level confidence intervals for $S_1,\ldots,S_p$ can be written as the
sum of two terms. The first, depending on $N$, accounts for sampling error and
can be modelled as
\[ \frac{Z_\alpha}{\sqrt{N}} \]
for a constant $Z_\alpha>0$. The assumption
of $1/\sqrt N$ decay is heuristically deduced from central limit theorem.

The second term, which accounts for metamodel error, is assumed to be of exponential decay when $n$ increases: $C/a^n$, where $C>0$ and $a>1$ are constants. This assumption is backed up by numerical experiments as well as theoretical works \cite{buffa2009apriori}.

Once this analysis has been done, we translate our problem into the following constrained minimization program:
\begin{equation}\label{e:coptprob} \text{Find } (N^*,n^*) = \argmin_{(n,N)\in\R^+\times\R^+} n^3 \times N \text{ so that }
\frac{2 q_\alpha \sigma}{\sqrt{N}} + \frac{C}{a^n} = P \end{equation}
where $P$ is the desired average precision for the confidence intervals.

Note that we converted the discrete design variables $N$ and $n$ to continuous positive variables so as to use the standard tools of continuous optimization; once optimum of the continuous problem have been found, we just round it to the nearest integer couple to recover a near-optimal integer solution.

\subsubsection*{Resolution of the optimization problem} 
The constraint in \eqref{e:coptprob} is equivalent to the conjunction of the following two equations:
\begin{gather}
\label{e:NN}
N = \left( \frac{Z_\alpha}{P-\frac{C}{a^n}} \right)^2 \\ 
\label{e:const}
P - \frac{C}{a^n} \geq 0 
\end{gather}

so that the function to minimize over $I = \left]n_{c};+\infty \right[$ (where $n_{c} = \ln(C/P)/\ln(a)$  has been chosed so as to satisfy \eqref{e:const}) is:
\[ \phi(n) = \left( Z_\alpha \right)^2 \frac{n^3}{\left( P - \frac{C}{a^n} \right)^2} \]
This function is differentiable on $I$ and tends to $+\infty$ as $n\rightarrow n_c$ and $n\rightarrow+\infty$; hence it has a minimizer $n^* \in I$ that satisfies $ \phi'(n^*)=0$, which is equivalent to:
\begin{equation}
\label{e:foc}
\frac{n^*}{P a^{n^*} - C} - \frac{3}{2 C \ln a} = 0
\end{equation}
On $I$, \eqref{e:foc} is
equivalent to:
\[ n^* - \frac{3P}{2C\ln a} a^{n^*} = - \frac{3}{2C\ln a} \]
Now let $\psi$ be the function defined on $\R$ by $\psi(x)=x-\frac{3 P}{2C\ln a}
a^x$. By remarking that $\psi'(x)=1-\frac{3P}{2C}a^x$ is continuous and nonzero
on $\left] \ln\left(\frac{2C}{3P}\right)/\ln a;+\infty \right[ \supset I$, one
has that $\psi$ is injective on $I$, and so \eqref{e:foc} has at most one
solution in $I$. Thus \eqref{e:foc} has exactly one solution in $I$, of which an
approximate value can be found by using bisection method
\cite{press1992numerical} on $[n_{c}+\epsilon ; L]$ where $\epsilon>0$ is small
enough and $L$ is a sufficiently large. Once $n^*$ has been found, we can 
find the optimal $N^*$ by setting $n=n^*$ in \eqref{e:NN}. 

\subsubsection*{Estimation of the parameters}
The last question that remains to address is the estimation of the $Z_\alpha$, $a$
and $C$ constants. The $Z_\alpha$ parameter is estimated by running the estimation
procedure on the metamodel, for fixed $N$ and $n$, estimating combined BC bootstrap confidence
intervals (Section \ref{s:bootstrap}) and taking:
\[ \widehat Z_\alpha=\sqrt N \left(\widehat S_{i,1-\alpha/2}^M-\widehat S_i^M + \widehat
S_{i,\alpha/2}^m - \widehat S_i^m\right) \]
where the factor multiplying $\sqrt N$ is the estimated ``Monte-Carlo part'' of
the error.

The $a$ and $C$ parameters are estimated by running an estimation procedure, for
a single $N$ fixed, and different reduced basis sizes $n_1, \ldots, n_K$, and
measuring, for each reduced basis size, the average metamodel error
$e(n_k)=\widehat S_i^M - \widehat S_i^m$, for $k=1,\ldots,K$. The $\{ (n_1,
e(n_1));\ldots;(n_K, e(n_K)) \}$ pairs are then used to fit the exponential
regression model $ e(n) = C/a^n $.

If one wants to estimate the sensitivity indices with respect to all variables
$i=1,\ldots,p$ for a single value of $N$ and $n$, one can use bootstrap
procedure to estimate Monte-Carlo errors $\widehat E_1,\ldots,\widehat E_p$  
for each of the $p$ sensitivity indices estimators:
	\[ \widehat E_i = \widehat S_{i,1-\alpha/2}^M-\widehat S_i^M + \widehat S_{i,\alpha/2}^m - \widehat S_i^m \]
and then to take:
\[ \widehat Z_\alpha = \frac{\sqrt N}{p} \sum_{i=1}^p \widehat E_i \]

\section{Numerical results and discussion}
In this section, we test our combined confidence interval procedure described earlier, and compare it with Monte-Carlo estimation on the full model (with bootstrap to assess sampling error), and with the procedure described in \cite{storlie2009implementation} and implemented in the \emph{CompModSA} R package. Our criteria of comparison are the CPU time needed to compute the intervals and the lengths of these intervals (the smaller the better).

In all our tests we take $\alpha=.05$ and $B=2000$
bootstrap replications. 

\subsection{Model set-up}
Let $u$, a function of space $x\in [0;1]$ (note that space variable $x$ is unrelated to input parameter vector $\xx$) and time $t\in[0,T]$ ($T>0$ is a fixed (i.e., known) parameter) satisfying the \emph{viscous Burgers' equation}:
\begin{equation}
\label{e:burg}
\frac{\partial u}{\partial t} + \frac{1}{2} \frac{\partial}{\partial x}(u^2) - \nu \frac{\partial^2 u}{\partial x^2} = \psi 
\end{equation}
where $\nu\in\R^+_*$ is the \emph{viscosity}, and $\psi \in C^0\left([0,T], L^2(]0,1[)\right)$ is the \emph{source term}.

For $u$ to be well-defined, we also prescribe initial value $u_0 \in H^1(]0,1[)$:
\begin{equation}
\label{e:init}
u(t=0,x)=u_0(x) \;\; \forall x\in[0;1]
\end{equation}
and boundary values $b_0, b_1 \in C^0([0,T])$:
\begin{equation}
\label{e:boundd}
\begin{cases} u(t,x=0)=b_0(t) \\ u(t,x=1)=b_1(t) \end{cases} \;\; \forall t\in[0;T]  
\end{equation}
Where $b_0$, $b_1$ and $u_0$ are given functions, supposed to satisfy \emph{compatibility conditions}:
\begin{equation}
\label{e:compatcond}
u_0(0)=b_0(0) \;\; \text{ and } \;\; u_0(1)=b_1(0) 
\end{equation}

The initial $u_0$ and boundary values $b_0$ and $b_1$ are parametrized the following way:
\begin{align*}
b_0(t) &= b_{0m} + \sum_{l=1}^{n(b_0)} A^{b_0}_l \sin( \omega^{b_0}_l t ) & 
b_1(t) &= b_{1m} + \sum_{l=1}^{n(b_1)} A^{b_1}_l \sin( \omega^{b_1}_l t ) \\
f(t,x) &= f_m + \sum_{l=1}^{n_T(f)} \sum_{p=1}^{n_S(f)} A^f_{lp} \sin( \omega^{fT}_l t ) \sin( \omega^{fS}_p x) &
u_0(x) &= \left(u_{0m}\right)^2 + \sum_{l=1}^{n(u_0)} A^{u_0}_l \sin( \omega^{u_0}_l x )
\end{align*}
The values of the angular frequencies $\omega^{b_0}_l$, $\omega^{b_1}_l$, $\omega^{fT}_l$, $\omega^{fS}_p$ and $\omega^{u_0}_l$, as well as their cardinalities $n(b_0)$, $n(b_1)$, $n_T(f)$, $n_S(f)$ and $n(u_0)$ are fixed (known), while our uncertain parameters, namely: viscosity $\nu$, coefficients $b_{0m}$, $b_{1m}$, $f_m$ and $u_{0m}$, and amplitudes $\left(A^{b_0}_l\right)_{l=1,\ldots,n(b_0)}$, $\left(A^{b_1}_l\right)_{l=1,\ldots,n(b_1)}$, $\left(A^f_{lp}\right)_{l=1,\ldots,n_T(f) ; p=1,\ldots,n_S(f)}$ and $\left(A^{u_0}_l\right)_{l=1,\ldots,n(u_0)}$ live in some Cartesian product of intervals $\mathcal P'$, subset of $\R^{1+4+n(b_0)+n(b_1)+n_T(f) n_S(f)+n(u_0)}$.

However, the compatibility condition \eqref{e:compatcond} constraints $b_{0m}$ and $b_{1m}$ as functions of the other parameters:
\begin{equation}
\label{e:compatcond2}
b_{0m} = \left( u_{0m} \right)^2 \;\;\text{ and }\;\; b_{1m}= \left(u_{0m}\right)^2+\sum_{l=1}^{n(u_0)} A_l^{u_0} \sin(\omega^{u_0}_l ) 
\end{equation}
so that the ``compliant'' uncertain parameters actually belong to $\mathcal P$ defined by:
\begin{multline}
\label{e:definitionP}
\mathcal P = \Big\{ \xx=\big(\nu,b_{0m},
							A^{b_0}_1,\ldots,A^{b_0}_{n(b_0)},
							b_{1m},A^{b_1}_1,\ldots,A^{b_0}_{n(b_1)}, 
							f_m, 
							A^f_{11},A^f_{12},\ldots,A^f_{1,n_S(f)}, \\
							A^f_{2,1},\ldots,A^f_{2,n_S(f)},\ldots,A^f_{n_T(f),n_S(f)}, 
							u_{0m},A^{u_0}_1,\ldots,A^{u_0}_{n(u_0)} \big) \in \mathcal P' 
							\text{ satisfying } \eqref{e:compatcond2} \Big\}
\end{multline}
In \cite{janon2011certified}, we gave an example with many more parameters. To illustrate our sensitivity analysis methodology without overloading the text, we choose an example with a reduced number of parameters.

The solution $u=u(\xx)$ depends on the parameter vector $\xx$ above.

The ``full'' model is obtained by discretizing the initial-boundary value problem \eqref{e:burg}, \eqref{e:init}, \eqref{e:boundd}, using a discrete time grid $\{ t_k=k \Delta t \}_{k=0,\ldots,T/\Delta t}$, where $\Delta t > 0$ is the time step, and, space-wise, using $\mathbf P^1$ Lagrange finite elements built upon an uniform subdivision of $[0;1]$: $\left\{ x_i = i / \mathcal N \right\}$, for $i=0,\ldots,\mathcal N$. Our full output is:
\[ f(\xx) = f(u(\xx)) = \frac{1}{\mathcal N} \sum_{i=0}^{\mathcal N} u(t=T, x=x_i) \]
The reduced basis method is then applied to yield a surrogate solution $\widetilde u$ of \eqref{e:burg}, \eqref{e:init}, \eqref{e:boundd}, as well as an error bound $\epsilon_u$ on $u$. Due to non-linearity and time-dependence of \eqref{e:burg}, as well as parametrization of the boundary values, the reduced basis methodology is not as simple as the one presented in Section \ref{s:RB} of the present paper. The reader can refer to \cite{janon2011certified} for full details on discretization and reduction of this model. Error bound on output $\epsilon$ is obtained by following \eqref{e:errc}.

In our numerical experiments, we take $\mathcal N=60$, $\Delta t=.01$, $T=.05$, 
$n_S(f)=n_T(f)=n(b_0)=n(b_1)=0$, $n(u_0)=1$, $\omega_1^{u_0}=0.5$, $A_{1}^{u_0}=5$ and $f_m=1$.

Reduced basis are found using POD-based procedure with $\#\Xi=30$.

\begin{figure}
\begin{center}
\includegraphics[width=10cm]{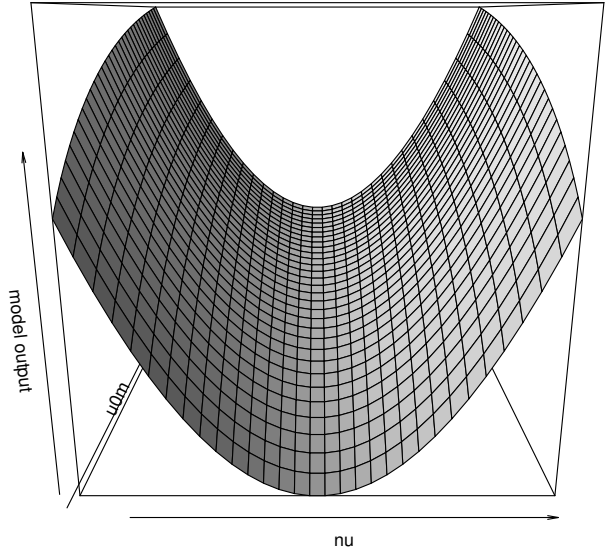}
\caption{Output $f$ of the Burgers' model, plotted as a function of $\nu$ and $u_{0m}$.}
\label{f:plot}
\end{center}
\end{figure}

The two input parameters are assumed independent and uniformly distributed. The
table below contains the ranges for them, and also the ``true'' values of the
sensitivity indices, which have been calculated (in more than 14h CPU time) using a
Monte-Carlo simulation with large sample size $N=4\times10^6$ (so as to BC
bootstrap confidence intervals of length $< 10^{-2}$) on the full model:
\begin{center}
{\footnotesize
\begin{tabular}{|c|c|c|}
\hline Parameter & Range & $95\%$ confidence interval for sensitivity index \\
\hline $\nu$ & [1 ; 20] & [0.0815;0.0832] \\
\hline $u_{0m}$ & [-0.3 ; 0.3] & [0.9175;0.9182] \\
\hline
\end{tabular} }
\end{center}
The output, as a function of the two uncertain parameters $\nu$ and $u_{0m}$ is plotted at Figure \ref{f:plot}; as one can see it is nonlinear with respect to the input parameters.

\subsection{Convergence benchmark}
\begin{figure}
\begin{center}
\includegraphics{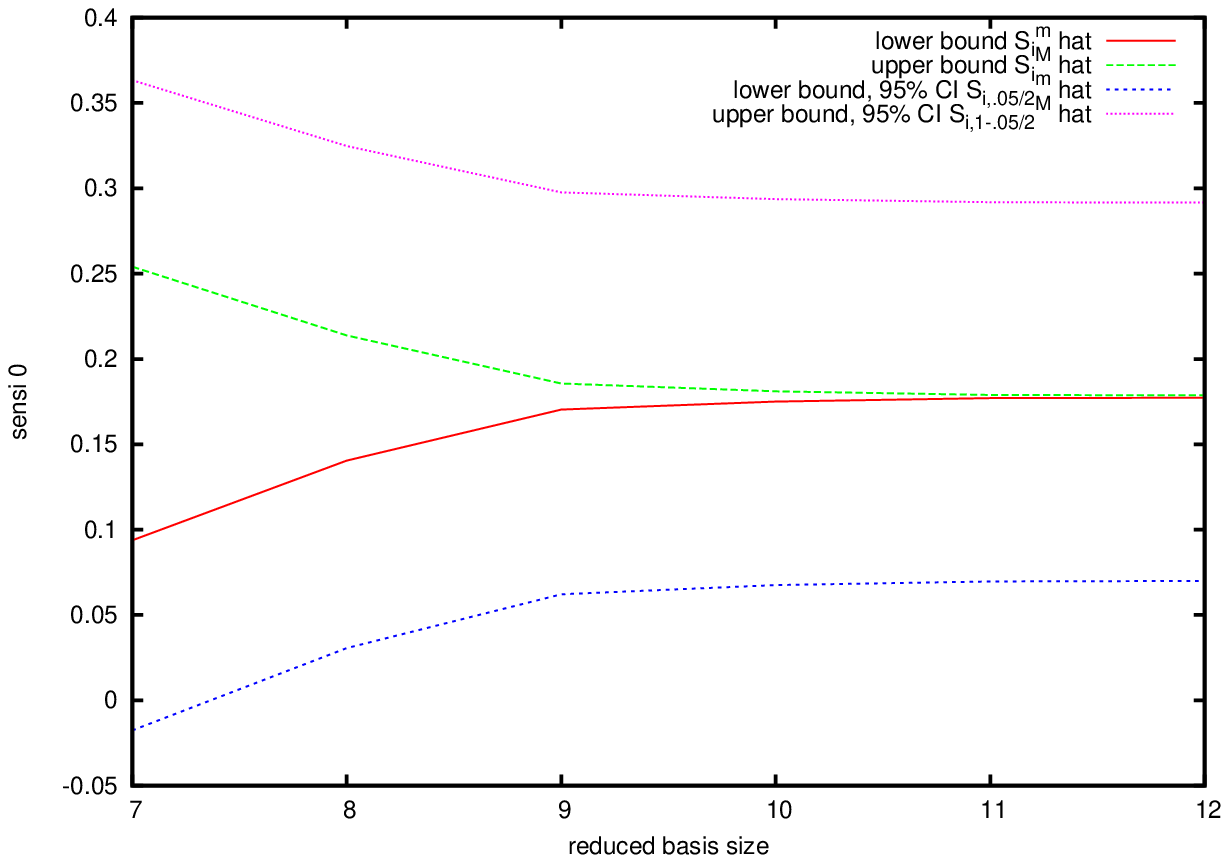} \\
\includegraphics{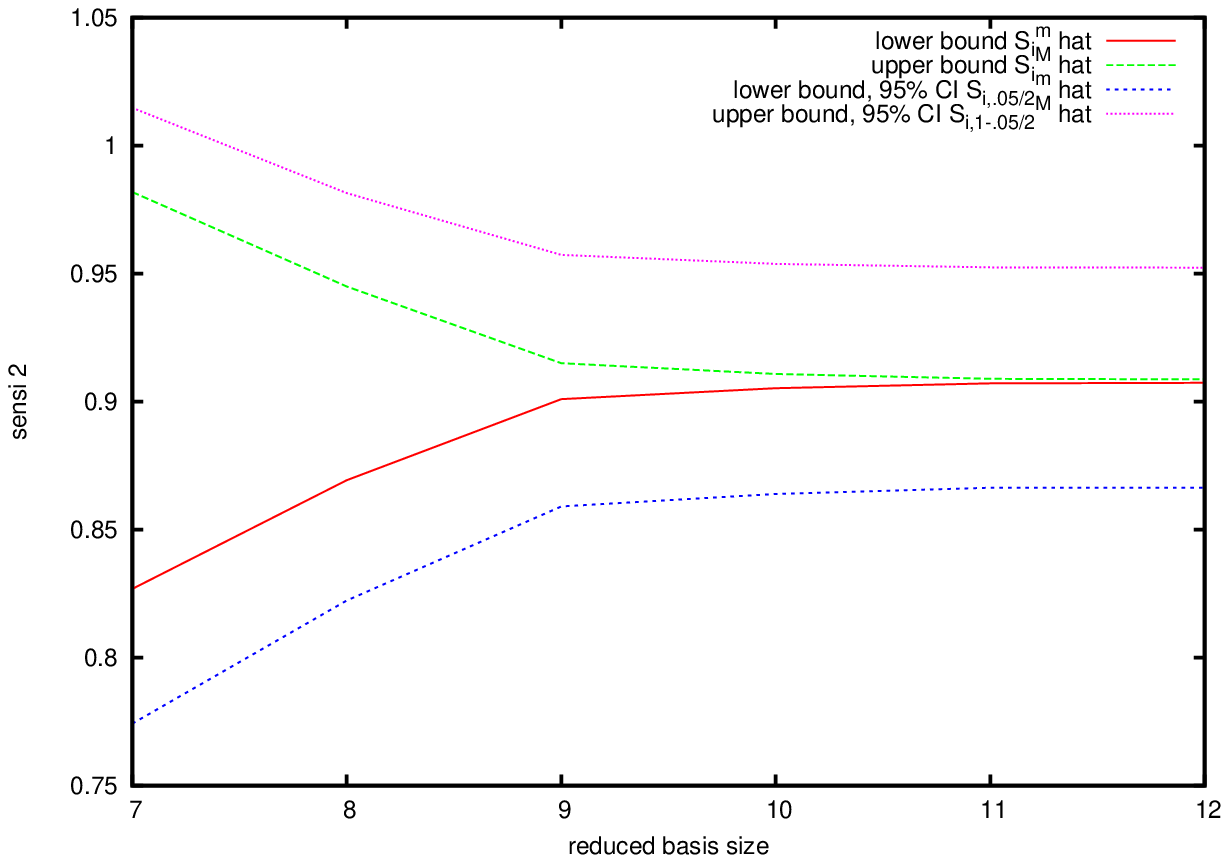} \\
\caption{Convergence benchmark for sensitivity indices estimation in the
Burgers' model, top: variable $\nu$, bottom: variable $u_{0m}$. We plotted, for
a fixed sample size $N=300$, estimator bounds $\widehat S^m$ and $\widehat S^M$
defined in \eqref{ss:mmerror}, and endpoints $\widehat S_{i,.025}^m$ and
$\widehat S_{i,1-.025}^M$ of the 95\% confidence interval, for different reduced basis sizes.}
\label{f:fig1}
\end{center}
\end{figure}

Figure \ref{f:fig1} shows the lower $\widehat{S^{m}}$ and upper
$\widehat{S^{M}}$ bound (defined in Section \ref{ss:mmerror}) for different
reduced basis sizes (hence different metamodel precision) but fixed sample of
size $N=300$, as well as the bootstrap confidence intervals computed using the procedure presented in Section \ref{s:bootCI}. This figure exhibits the fast convergence of our bounds to the true value of the sensitivity index as the reduced basis size increases. We also see that the part of the error due to sampling (gaps between confidence interval upper bound and upper bound, and between confidence interval lower bound and lower bound) remains constant, as sample size stays the same.

\subsection{Choice of $n$ and $N$}
\label{ss:choiceNn}
\begin{figure}
\begin{center}
\includegraphics{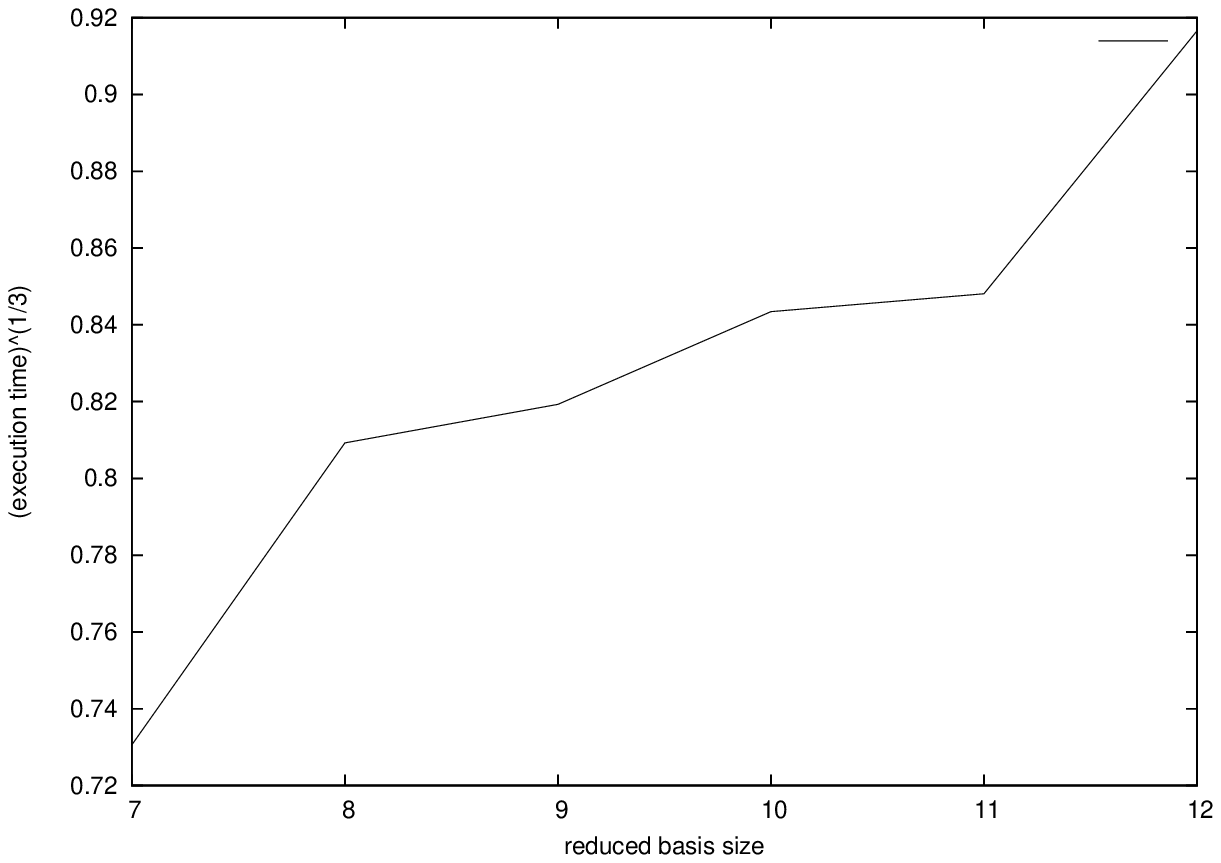} 
\caption{Cube root of the CPU time necessary to do estimations, as a function of
the reduced basis size of $n$. Section \ref{s:towards} assumes this function is linear. }
\label{f:fig2}
\end{center}
\end{figure}

\begin{figure}
\begin{center}
\includegraphics{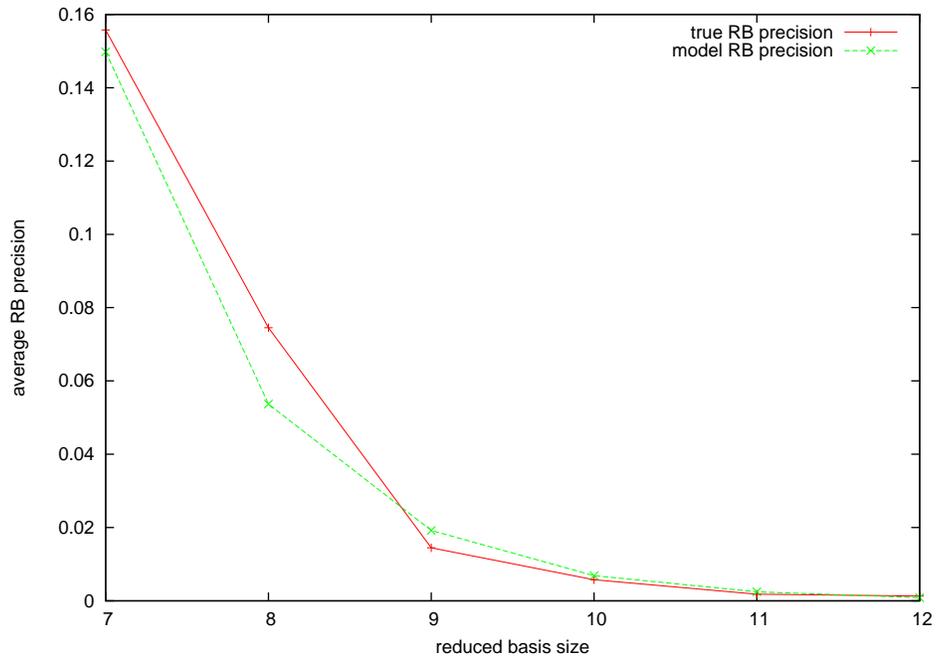} 
\caption{Line: reduced basis ``precision'', i.e. mean $\widehat S_i^M-\widehat S_i^m$, as a function of reduced basis size; dashes: result of the fit of an exponential regression model: $\widehat C/\widehat a^n$. }
\label{f:fig3}
\end{center}
\end{figure}

We now discuss the numerical results obtained when using the parameter tuning
procedure (Section \ref{s:towards}). We have done ``pre-runs'' for $N=300$, and different reduced basis sizes $\{7,8,\ldots,12\}$ of the combined confidence interval procedure, to yield the following estimates:
\[ \widehat C=197.69 \;\;\; \widehat a = 2.789 \;\;\; \widehat Z_{.05}=2.6407 \]
To assess validity of this estimation, and to check our modelisation of the
bound precision and the execution time, we plot the cube root of the CPU time
(Figure \eqref{f:fig2}) and the precision of the bound defined in Section \ref{ss:mmerror} (Figure \eqref{f:fig3}). We can check that these hypotheses are reasonably satisfied.

One can find in Figure \ref{f:4} the computed optimal reduced basis sizes $n^*$ and sample sizes $N^*$ using resolution of the optimization problem \eqref{e:coptprob}, for various precisions $p$.

\begin{figure}
\begin{center}
\begin{tabular}{|c|c|c|}\hline Precision $p$ & Reduced basis size $n^*$ & Sample size $N^*$ \\
\hline 0.005&12.4437&354491\\
\hline 0.02&11.1095&22057.6\\
\hline 0.05&10.0501&3698.95\\
\hline 0.08&9.59689&1442.7\\
\hline 0.09&9.48332&1139.47\\
\hline
\end{tabular}
\caption{Optimal reduced basis and sample sizes calculated using the strategy
described in Section \ref{s:towards}. }
\label{f:4}
\end{center}
\end{figure}
All these values have been computed in 5.77 s CPU time, including the time necessary to estimate $C$, $a$ and $\sigma$.

To check for the efficiency of this parameter tuning strategy, we choose a target precision of $p=.02$. In the table \ref{f:4}, we read that we should take $n \approx 11$ and $N \approx 22000$.

Conducting the combined confidence interval estimation with these parameters
give intervals \\
$[ 0.0659997 ; 0.0937285 ]$ for sensitivity index for $\nu$, and $[ 0.914266 ; 0.926452 ]$ for sensitivity with respect to $u_{0m}$. These confidence intervals have mean length:
\[ \frac{1}{2} \left(0.0937285 - 0.0659997 + 0.926452 - 0.914266 \right) =
0.0199575 \approx 0.02 \]
as desired.

This computation took 52 s of CPU time to complete (including a metamodel offline phase of 1 s).

\subsection{Normality of the bootstrap distributions}
\label{ss:normapprox}
\begin{figure}
\begin{center}
\includegraphics[angle=-90,width=6cm]{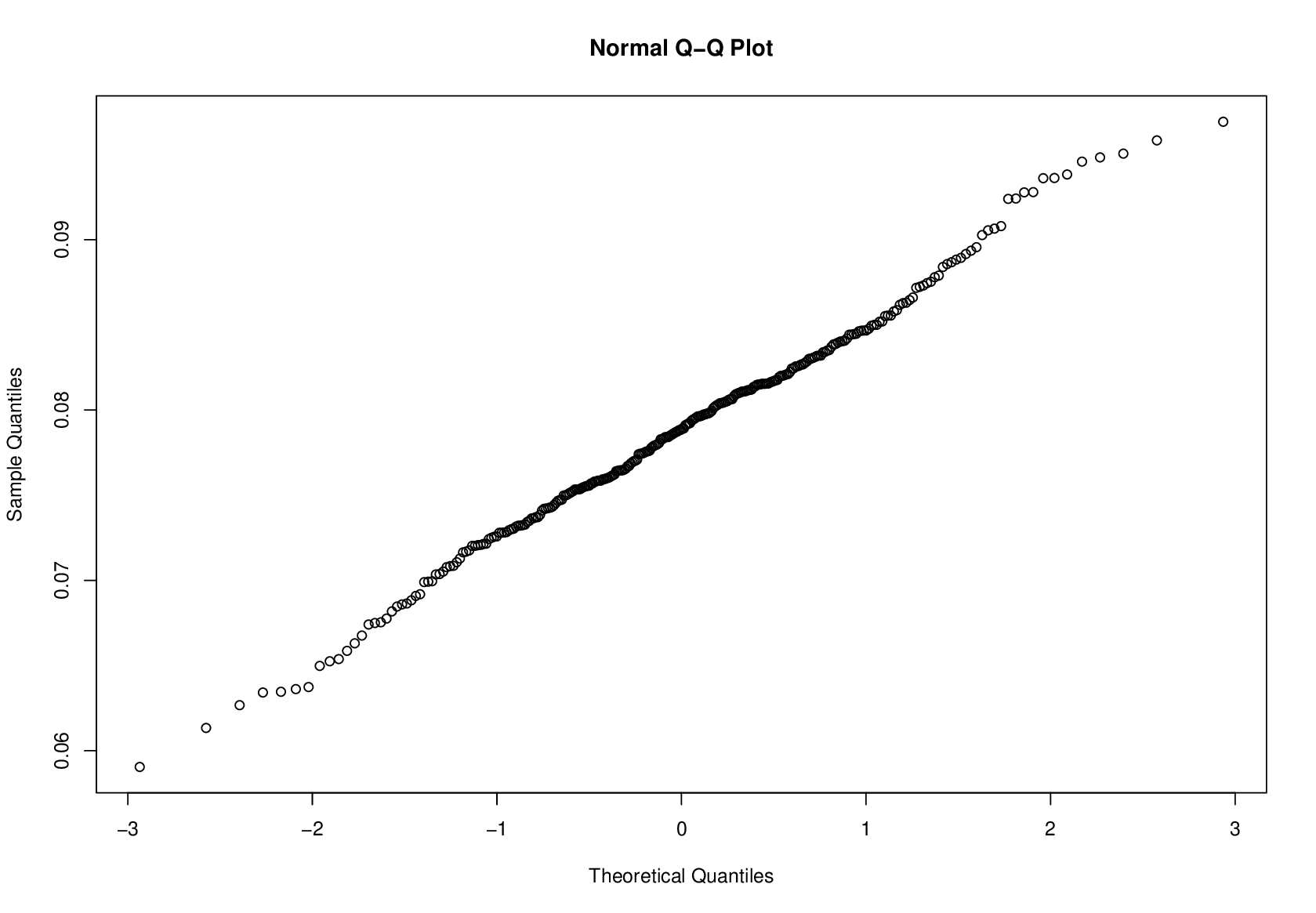}
\includegraphics[angle=-90,width=6cm]{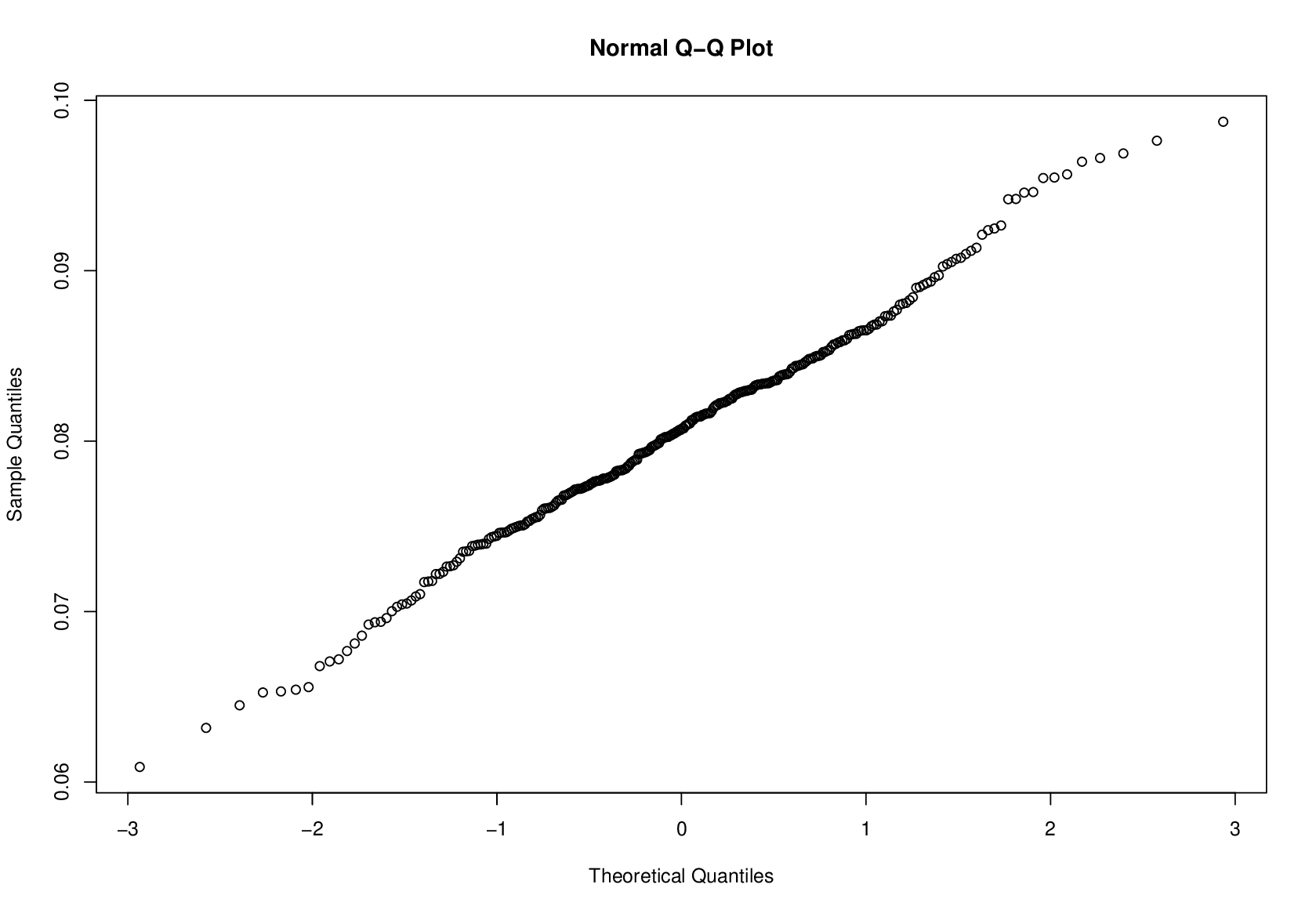}
\caption{Normal empirical quantile-quantile plots of the distributions of
$S_\nu^m$ (left) and $S_\nu^M$ (right). }
\label{f:qqplot}
\end{center}
\end{figure}

We give in Figure \ref{f:qqplot} the empirical normal quantile-quantile plots of the bootstrap replicates 
$\{ \widehat S_i^m[1], \ldots, \widehat S_i^m[B] \}$  and  $\{ \widehat S_i^M[1],\ldots, \widehat
S_i^M[B] \}$.

As these plots are close to a line, the bootstrap distributions are approximately normal.

\subsection{Optimality of our metamodel error bound}
We checked for near optimality of the metamodel error bound \ref{ss:mmerror} by
comparing it with the values of the optimization problems: $\min_{\mathcal
Y} \psi$ and $\max_{\mathcal Y}\psi$ where:
\[ \psi(\yy)= \frac{\frac{1}{N} \sum_{k=1}^N y_k y_k' - \left( \frac{1}{N} \sum_{k=1}^N y_k \right) \left( \frac{1}{N} \sum_{k=1}^N y_k' \right) }{  \frac{1}{N} \sum_{k=1}^N \left( y_k \right)^2 - \left( \frac{1}{N} \sum_{k=1}^N y_k \right)^2 } \]
and:
\[ \mathcal Y = \prod_{k=1}^N \left[ \widetilde y_k - \epsilon_k ; \widetilde y_k + \epsilon_k \right]
\times
\prod_{k=1}^N \left[ \widetilde y_k' - \epsilon_k' ; \widetilde y_k' + \epsilon_k' \right] \] 
These problems, of large dimension $2N$, give the optimal values of $\widehat S_i^m$ and $\widehat S_i^M$
satisfying \eqref{e:boundSI}. They can be solved with simulated annealing
\cite{kirkpatrick1983optimization,pardalos2002handbook}.

Our bound gave results very close to the optimal ones, for a smaller computational cost
than using simulated annealing.

\subsection{Comparison with estimation on the full model}
To obtain a result of the same precision, we carry a simulation for $N=21000$ (sample size can be chosen smaller than before, as there will be no metamodel error) on the \emph{full} model; we get the bootstrap confidence interval with mean length of $\approx0.0193$.

This computation takes 308 s of CPU time to complete. Hence, on this example, using a reduced-basis surrogate model roughly divides overall computation time by a factor of 5.9, without any sacrifice on the precision and the rigorousness (as our metamodel error quantification procedure is fully proven and certified) of the confidence interval. We expect higher time savings with more complex (for example, two- or three-dimensional in space) models. 
 
\subsection{Comparison with CompModSA}
We compared our results with the ones obtained using the R \emph{CompModSA}
version 1.2 package downloaded at \cite{compmodsa}. This package implements the
method described in \cite{storlie2009implementation} for assessing sampling
error as well as metamodel error. It does not make use of the reduced basis
output error bound, but uses a non-intrusive method to fit a metamodel using a
reasonable number of full model evaluations.

We fed into CompModSA procedure 50 such full model outputs (which took 0.22 s
CPU to compute). We then tried various non-intrusive metamodels (\emph{surface
responses}), and reported the results into Figure \ref{f:5}. We used as
parameters: n.mc.T=0 (we do not want any total index computation), n.mc.S=23000 (sample size), n.samples=1 (one run), n.CI=300 (generate confidence intervals using 300 bootstrap replications). We contributed \footnote{patch available at \url{http://ljk.imag.fr/membres/Alexandre.Janon/compmodsa.php}} the option CI.S, which is set to TRUE to compute bootstrap confidence intervals for the main effect index.

\begin{figure}
\begin{center}
	{\small
\begin{tabular}{|c|c|c|c|}\hline Surface response (metamodel) & Mean confidence interval length & $R^2$ & CPU time \\
\hline qreg: quadratic regression & 0.081 & 0.996799 & 143.59 s \\
\hline mars: multivariate adaptive regression splines & 0.075 & 0.9998506 & 218.716 s \\
\hline \emph{our approach} & 0.019 & N/A & 52 s \\
\hline
\end{tabular}
\caption{Results of CompModSA's sensitivity function on our model, for two
fitted response surfaces. $R^2$ is an indicator of the metamodel fit (values
close to unity suggest good fit). The last line recalls the results of the
experiment in Section \ref{ss:choiceNn}. }
\label{f:5} }
\end{center}
\end{figure}

By looking at results in Figure \ref{f:4}, we can see that in
this case, our approach is clearly superior, both in terms of precision and
computation time. To achieve this result, we took advantage of the
particular formulation of the original model which allows the reduced basis
methodology to be efficiently applied; CompModSA, due to its non-intrusive nature, is easier
to use on a generic ``black box'' model.




\section*{Conclusion and perspectives}
We presented a methodology to make a rigorous quantification of the impact of
the sampling error and the metamodel error on the sensitivity indices
computation, when a reduced-basis metamodel is used. Sampling error is handled
by a classic bootstrap procedure, while metamodel error is managed using a new
bound on the sensitivity index estimator. Quantification of those two types of
errors permits not only a certification on the performed estimation, but also
gives a way to tune the optimal parameters (reduced basis and optimal sample
sizes), for a given desired precision on the indices. We have shown on a
concrete example the superiority of this method when compared to the use of
the full model, or non-intrusive (quadratic regression, MARS) metamodels. Our
method can be applied to other Monte Carlo (or quasi Monte Carlo) estimators,
and to other metamodels which provide an error bound similar to the one provided
by the reduced basis framework. 

\paragraph{Acknowledgements} We wish to thank Jean-Claude Fort for a
suggestion which we exploited to perform our computation of the
metamodel-induced error. We also thank Anestis Antoniadis and Ingrid Van
Keilegom for advice on bootstrap methodology. -- This work has been partially supported by the French National
Research Agency (ANR) through COSINUS program (project COSTA-BRAVA
n°ANR-09-COSI-015).

\setcounter{section}{0}
\renewcommand{\thesection}{\Alph{section}}

\section{Algorithms}
\label{appendix:B}
\begin{algo}
\label{a:bootmc}
\begin{enumerate}
\item Draw from $\XX$ distribution two independent samples of size $N$:
$\{\xx^k\}$ and $\{\xx'^k\}$.
\item Tabulate necessary model evaluations: for $k=1,\ldots,N$:
 \begin{enumerate}
 \item set $\xx \leftarrow \xx^k$;
 \item compute $y_k = f(\xx)$;
 \item swap $X_i$ and $X'^k_i$;
 \item compute $y_k'=f(\xx)$;
 \item swap $X_i$ and $X'^k_i$ back.
 \end{enumerate}
\item Compute $\widehat S_i$:
 \[ \widehat S_i = \frac{\frac{1}{N} \sum_{k=1}^N y_k y_k' - \left(
 \frac{1}{N} \sum_{k=1}^N y_k \right) \left( \frac{1}{N} \sum_{k=1}^N y_k'
 \right) }{  \frac{1}{N} \sum_{k=1}^N y_k^2 - \left( \frac{1}{N} \sum_{k=1}^N y_k \right)^2 } \] 
\item Repeat, for $b=1,\ldots,B$:
 \begin{enumerate}
 \item Draw at random a list $L$ of length $N$, with replacement from $\{1,\ldots,N\}$.
 \item Compute replication $\widehat S_i [b]$ :
 \[ \widehat S_i [b] = \frac{\frac{1}{N} \sum_{k\in L} y_k y_k' - \left(
 \frac{1}{N} \sum_{k \in L} y_k \right) \left( \frac{1}{N} \sum_{k \in L} y_k'
 \right) }{  \frac{1}{N} \sum_{k \in L} y_k^2 - \left( \frac{1}{N} \sum_{k \in
 L} y_k \right)^2 } \] 
 \end{enumerate}
\item Compute $\widehat z_0$:
\[ \widehat{z_0} = \Phi^{-1}\left( \frac{ \#\{b\in\{1,\ldots,B\} \textrm{~s.t.~}
\widehat S_i[b]\leq\widehat S_i \} }{B} \right) \]
where $\Phi(z) = \frac{1}{\sqrt{2\pi}} \int_{-\infty}^z \exp\left(- \frac{t^2}{2}
\right) \ud t$.
\item Look up for $z_{\alpha/2}$ so that:
\[ \Phi(z_{\alpha/2})=\alpha/2 \]
and take $z_{1-\alpha/2}=-z_{\alpha/2}$, satisfying:
$ \Phi(z_{1-\alpha/2})=1-\alpha/2 $. 
\item Compute $\widehat q(\alpha/2)$ and $\widehat q(1-\alpha/2)$:
\[ \widehat q(\alpha/2) = \Phi(2\widehat{z_0}+z_{\alpha/2}), \;\;\;\;\;
\widehat q(1-\alpha/2) = \Phi(2\widehat{z_0}+z_{1-\alpha/2}) \]
\item Compute $\widehat S_{i,\alpha/2}$ and $\widehat S_{i,1-\alpha/2}$, the $
\widehat q(\alpha/2)$ and $\widehat q(1-\alpha/2)$ quantiles of $\{
	S_i[1],\ldots,S_i[B] \}$.
\item Output $\left[ \widehat S_{i,\alpha/2} ; \widehat S_{i,1-\alpha/2}
\right]$ as confidence interval for $S_i$ of level $1-\alpha$.

\end{enumerate}
\end{algo}

\begin{algo}
\label{a:mcerr}
\begin{enumerate}
\item Draw from $\XX$ distribution two independent samples of size $N$:
$\{\xx^k\}$ and $\{\xx'^k\}$.
\item Tabulate necessary model evaluations: for $k=1,\ldots,N$:
 \begin{enumerate}
 \item set $\xx \leftarrow \xx^k$;
 \item compute $\widetilde y_k = \widetilde f(\xx)$ and $\epsilon_k = \epsilon(\xx)$ ;
 \item swap $X_i$ and $X'^k_i$;
 \item compute $\widetilde y_k' = \widetilde f(\xx)$ and $\epsilon_k' = \epsilon(\xx)$;
 \item swap $X_i$ and $X'^k_i$ back.
 \end{enumerate}
\item Compute $\overline{\widetilde y}, \overline{\widetilde y'},
\overline{\epsilon}$ and $\overline{\epsilon'}$, the respective means of
$\{\widetilde y_k\}, \{\widetilde y_k'\}, \{\epsilon_k\}$ and $\{\epsilon_k'\}$.
\item Choose a finite subset $\Xi$ of the set $\overline{\mathcal P} \times
\overline{\mathcal P'}$ where $\overline{\mathcal P}$ and $\overline{\mathcal
P'}$ are defined by \eqref{e:defp} and \eqref{e:defppr}.
\item Repeat, for $(\mu,\mu')\in\Xi$:
 \begin{enumerate}
 \item By using \eqref{e:defrinf} and \eqref{e:defrsup}, compute
 $R_{inf}(a;\widetilde{\yy},\epsilon,\mu,\mu')$ and
 $R_{sup}(a;\widetilde{\yy},\epsilon,\mu,\mu')$ for three different values of
 $a$; 
 \item deduce $\alpha_{inf}, \beta_{inf}, \gamma_{inf}, \alpha_{sup},
 \beta_{sup}$ and $\gamma_{sup}$ satisfying \eqref{e:coefrinf} and
 \eqref{e:coefrsup};
 \item if $\alpha_{inf} \leq 0$, exit with failure;
 \item compute
 $\widehat\delta=2\sqrt{(\alpha_{inf}-\alpha_{sup})(\gamma_{inf}-\gamma_{sup})}$;
 \item compute:
 \[ \widehat S_i^m(\mu,\mu') = - \frac{\beta_{inf} + \widehat\delta}{2 \alpha_{inf}}
 \;\;\;\;\;\;
  \widehat S_i^M(\mu,\mu') = - \frac{\beta_{sup} - \widehat\delta}{2 \alpha_{sup}}
 \]
 \end{enumerate}
\item Output:
\[ \widehat S_i^m = \min_{(\mu,\mu')\in\Xi} \widehat S_i^m(\mu,\mu')
\;\;\;\;\;\;
\widehat S_i^M = \max_{(\mu,\mu')\in\Xi} \widehat S_i^M(\mu,\mu')
\]
\end{enumerate}
\end{algo}

\begin{algo}
\label{a:confInt}
\begin{enumerate}
\item Follow steps 1. and 2. of Algorithm \ref{a:mcerr}.
\item Compute bounds  $\widehat S_i^m$ and $\widehat S_i^M$ using steps
 3.--6. of Algorithm \ref{a:mcerr}.
\item Repeat, for $b=1,\ldots,B$:
 \begin{enumerate}
 \item Draw at random a list $L$ of length $N$, with replacement from $\{1,\ldots,N\}$.
 \item If using $\epsilon$-sampling: for $k\in L$, sample $\epsilon_k^*$ uniformly
	 in $\left[\eta_k \epsilon_k ; \epsilon_k\right]$ and ${\epsilon'}_k^*$ uniformly in
 $\left[\eta_k' \epsilon_k' ; \epsilon_k'\right]$ .
 \item Else: take, for $k\in L$,  $\epsilon_k^*=\epsilon_k$ and
	 ${\epsilon'}_k^*=\epsilon_k'$.
 \item Compute bounds $\widehat S_i^m[b]$ and $\widehat S_i^M[b]$ using steps
 3.--6. of Algorithm \ref{a:mcerr}, with $\left(\widetilde y_k\right)_{k\in L}$ instead of
 $\left(\widetilde y_k\right)_{k=1,\ldots,N}$, $\left(\widetilde y_k'\right)_{k\in L}$ instead of
 $\left(\widetilde y_k'\right)_{k=1,\ldots,N}$ as sample data, and
 $\left(\epsilon_k^*\right)_{k\in L}$ instead of
 $\left(\epsilon_k\right)_{k=1,\ldots,N}$, and $\left({\epsilon'}_k^*\right)_{k\in L}$ instead of
 $\left(\epsilon_k'\right)_{k=1,\ldots,N}$ as error bounds.
 \end{enumerate}
\item Compute, for $w\in\{m,M\}$, the two bias correction constants:
\[ \widehat{z_0^w} = \Phi^{-1}\left( \frac{ \#\{b\in\{1,\ldots,B\} \textrm{~s.t.~}
\widehat S_i^w[b]\leq\widehat S_i^w \} }{B} \right) \]
where $\Phi(z) = \frac{1}{\sqrt{2\pi}} \int_{-\infty}^z \exp\left(- \frac{t^2}{2}
\right) \ud t$.
\item Look up for $z_{\alpha/2}$ so that:
\[ \Phi(z_{\alpha/2})=\alpha/2 \]
and take $z_{1-\alpha/2}=-z_{\alpha/2}$, satisfying:
$ \Phi(z_{1-\alpha/2})=1-\alpha/2 $. 
\item Compute $\widehat q^m(\alpha/2)$ and $\widehat q^M(1-\alpha/2)$:
\[ \widehat q^m(\alpha/2) = \Phi(2\widehat{z_0^m}+z_{\alpha/2}), \;\;\;\;\;
\widehat q^M(1-\alpha/2) = \Phi(2\widehat{z_0^M}+z_{1-\alpha/2}) \]
\item Compute $\widehat S_{i,\alpha/2}^m$ and $\widehat
S_{i,1-\alpha/2}^M$, the $
\widehat q^m(\alpha/2)$ and $\widehat q^M(1-\alpha/2)$ quantiles of $\{
	S_i^m[1],\ldots,S_i^m[B] \}$ and $\{
	S_i^M[1],\ldots,S_i^M[B] \}$, respectively.
\item Output $\left[ \widehat S_{i,\alpha/2}^m ; \widehat S_{i,1-\alpha/2}^M
\right]$ as combined confidence interval for $S_i$ of level $1-\alpha$.

\end{enumerate}
\end{algo}

\bibliographystyle{model2-names}
\bibliography{biblio}

\begin{thebibliography}{25}
\expandafter\ifx\csname natexlab\endcsname\relax\def\natexlab#1{#1}\fi
\expandafter\ifx\csname url\endcsname\relax
  \def\url#1{\texttt{#1}}\fi
\expandafter\ifx\csname urlprefix\endcsname\relax\def\urlprefix{URL }\fi
\providecommand{\eprint}[2][]{\url{#2}}
\providecommand{\bibinfo}[2]{#2}
\ifx\xfnm\relax \def\xfnm[#1]{\unskip,\space#1}\fi
\bibitem[{Archer et~al.(1997)Archer, Saltelli and
  Sobol}]{archer1997sensitivity}
\bibinfo{author}{Archer, G.}, \bibinfo{author}{Saltelli, A.},
  \bibinfo{author}{Sobol, I.}, \bibinfo{year}{1997}.
\newblock \bibinfo{title}{{Sensitivity measures, ANOVA-like techniques and the
  use of bootstrap}}.
\newblock \bibinfo{journal}{Journal of Statistical Computation and Simulation}
  \bibinfo{volume}{58}, \bibinfo{pages}{99--120}.
\bibitem[{Bergmann and Iollo(2008)}]{bergmann2008numerical}
\bibinfo{author}{Bergmann, J.}, \bibinfo{author}{Iollo, A.},
  \bibinfo{year}{2008}.
\newblock \bibinfo{title}{{Numerical methods for low-order modeling of fluid
  flows based on POD}} .
\bibitem[{Boyaval et~al.(2009)Boyaval, Bris, Maday, Nguyen and
  Patera}]{boyaval2009reduced}
\bibinfo{author}{Boyaval, S.}, \bibinfo{author}{Bris, C.},
  \bibinfo{author}{Maday, Y.}, \bibinfo{author}{Nguyen, N.},
  \bibinfo{author}{Patera, A.}, \bibinfo{year}{2009}.
\newblock \bibinfo{title}{{A reduced basis approach for variational problems
  with stochastic parameters: Application to heat conduction with variable
  robin coefficient}}.
\newblock \bibinfo{journal}{Computer Methods in Applied Mechanics and
  Engineering} \bibinfo{volume}{198}, \bibinfo{pages}{3187--3206}.
\bibitem[{Buffa et~al.(2009)Buffa, Maday, Patera, Prud'homme and
  G.}]{buffa2009apriori}
\bibinfo{author}{Buffa, A.}, \bibinfo{author}{Maday, Y.},
  \bibinfo{author}{Patera, A.}, \bibinfo{author}{Prud'homme, C.},
  \bibinfo{author}{G., T.}, \bibinfo{year}{2009}.
\newblock \bibinfo{title}{{A priori convergence of the greedy algorithm for the
  parametrized reduced basis}}.
\newblock \bibinfo{journal}{Mathematical Modelling and Numerical Analysis} .
\bibitem[{Bui-Thanh et~al.(2007)Bui-Thanh, Willcox, Ghattas and van
  Bloemen~Waanders}]{bui2007goal}
\bibinfo{author}{Bui-Thanh, T.}, \bibinfo{author}{Willcox, K.},
  \bibinfo{author}{Ghattas, O.}, \bibinfo{author}{van Bloemen~Waanders, B.},
  \bibinfo{year}{2007}.
\newblock \bibinfo{title}{{Goal-oriented, model-constrained optimization for
  reduction of large-scale systems}}.
\newblock \bibinfo{journal}{Journal of Computational Physics}
  \bibinfo{volume}{224}, \bibinfo{pages}{880--896}.
\bibitem[{Chatterjee(2000)}]{chatterjee2000introduction}
\bibinfo{author}{Chatterjee, A.}, \bibinfo{year}{2000}.
\newblock \bibinfo{title}{{An introduction to the proper orthogonal
  decomposition}}.
\newblock \bibinfo{journal}{Current Science} \bibinfo{volume}{78},
  \bibinfo{pages}{808--817}.
\bibitem[{compmodsa()}]{compmodsa}
compmodsa, .
\newblock \bibinfo{title}{{CompModSA}}.
\newblock
  \bibinfo{howpublished}{\url{http://www.stat.unm.edu/~storlie/CompModSA/}}.
\bibitem[{Efron(1981)}]{efron1981nonparametric}
\bibinfo{author}{Efron, B.}, \bibinfo{year}{1981}.
\newblock \bibinfo{title}{{Nonparametric standard errors and confidence
  intervals}}.
\newblock \bibinfo{journal}{Canadian Journal of Statistics}
  \bibinfo{volume}{9}, \bibinfo{pages}{139--158}.
\bibitem[{Efron and Tibshirani(1986)}]{efron1986bootstrap}
\bibinfo{author}{Efron, B.}, \bibinfo{author}{Tibshirani, R.},
  \bibinfo{year}{1986}.
\newblock \bibinfo{title}{{Bootstrap methods for standard errors, confidence
  intervals, and other measures of statistical accuracy}}.
\newblock \bibinfo{journal}{Statistical science} \bibinfo{volume}{1},
  \bibinfo{pages}{54--75}.
\bibitem[{Efron et~al.(1993)Efron, Tibshirani and
  Tibshirani}]{efron1993introduction}
\bibinfo{author}{Efron, B.}, \bibinfo{author}{Tibshirani, R.},
  \bibinfo{author}{Tibshirani, R.}, \bibinfo{year}{1993}.
\newblock \bibinfo{title}{{An introduction to the bootstrap}}.
\newblock \bibinfo{publisher}{Chapman \& Hall/CRC}.
\bibitem[{Grepl et~al.(2007)Grepl, Maday, Nguyen and
  Patera}]{grepl2007efficient}
\bibinfo{author}{Grepl, M.}, \bibinfo{author}{Maday, Y.},
  \bibinfo{author}{Nguyen, N.}, \bibinfo{author}{Patera, A.},
  \bibinfo{year}{2007}.
\newblock \bibinfo{title}{{Efficient reduced-basis treatment of nonaffine and
  nonlinear partial differential equations}}.
\newblock \bibinfo{journal}{Mathematical Modelling and Numerical Analysis}
  \bibinfo{volume}{41}, \bibinfo{pages}{575--605}.
\bibitem[{Grepl and Patera(2005)}]{grepl2005posteriori}
\bibinfo{author}{Grepl, M.}, \bibinfo{author}{Patera, A.},
  \bibinfo{year}{2005}.
\newblock \bibinfo{title}{{A posteriori error bounds for reduced-basis
  approximations of parametrized parabolic partial differential equations}}.
\newblock \bibinfo{journal}{Mathematical Modelling and Numerical Analysis}
  \bibinfo{volume}{39}, \bibinfo{pages}{157--181}.
\bibitem[{Helton et~al.(2006)Helton, Johnson, Sallaberry and
  Storlie}]{helton2006survey}
\bibinfo{author}{Helton, J.}, \bibinfo{author}{Johnson, J.},
  \bibinfo{author}{Sallaberry, C.}, \bibinfo{author}{Storlie, C.},
  \bibinfo{year}{2006}.
\newblock \bibinfo{title}{{Survey of sampling-based methods for uncertainty and
  sensitivity analysis}}.
\newblock \bibinfo{journal}{Reliability Engineering \& System Safety}
  \bibinfo{volume}{91}, \bibinfo{pages}{1175--1209}.
\bibitem[{Homma and Saltelli(1996)}]{homma1996importance}
\bibinfo{author}{Homma, T.}, \bibinfo{author}{Saltelli, A.},
  \bibinfo{year}{1996}.
\newblock \bibinfo{title}{{Importance measures in global sensitivity analysis
  of nonlinear models}}.
\newblock \bibinfo{journal}{Reliability Engineering \& System Safety}
  \bibinfo{volume}{52}, \bibinfo{pages}{1--17}.
\bibitem[{Janon et~al.(2010, \emph{submitted}.)Janon, Nodet and
  Prieur}]{janon2011certified}
\bibinfo{author}{Janon, A.}, \bibinfo{author}{Nodet, M.},
  \bibinfo{author}{Prieur, C.}, \bibinfo{year}{2010, \emph{submitted}.}
\newblock \bibinfo{title}{{Certified reduced-basis solutions of viscous Burgers
  equations parametrized by initial and boundary values}}.
\newblock \bibinfo{howpublished}{Preprint available at
  \url{http://hal.inria.fr/inria-00524727/en}}.
\bibitem[{Kirkpatrick et~al.(1983)Kirkpatrick, Gelatt~Jr and
  Vecchi}]{kirkpatrick1983optimization}
\bibinfo{author}{Kirkpatrick, S.}, \bibinfo{author}{Gelatt~Jr, C.},
  \bibinfo{author}{Vecchi, M.}, \bibinfo{year}{1983}.
\newblock \bibinfo{title}{{Optimization by simulated annealing}}.
\newblock \bibinfo{journal}{Science} \bibinfo{volume}{220},
  \bibinfo{pages}{671}.
\bibitem[{Marrel et~al.(2009)Marrel, Iooss, Laurent and
  Roustant}]{marrel2009calculations}
\bibinfo{author}{Marrel, A.}, \bibinfo{author}{Iooss, B.},
  \bibinfo{author}{Laurent, B.}, \bibinfo{author}{Roustant, O.},
  \bibinfo{year}{2009}.
\newblock \bibinfo{title}{{Calculations of sobol indices for the gaussian
  process metamodel}}.
\newblock \bibinfo{journal}{Reliability Engineering \& System Safety}
  \bibinfo{volume}{94}, \bibinfo{pages}{742--751}.
\bibitem[{Nguyen et~al.(2005)Nguyen, Veroy and Patera}]{nguyen2005certified}
\bibinfo{author}{Nguyen, N.}, \bibinfo{author}{Veroy, K.},
  \bibinfo{author}{Patera, A.}, \bibinfo{year}{2005}.
\newblock \bibinfo{title}{{Certified real-time solution of parametrized partial
  differential equations}}.
\newblock \bibinfo{journal}{Handbook of Materials Modeling} ,
  \bibinfo{pages}{1523--1558}.
\bibitem[{Pardalos and Romeijn(2002)}]{pardalos2002handbook}
\bibinfo{author}{Pardalos, P.}, \bibinfo{author}{Romeijn, H.},
  \bibinfo{year}{2002}.
\newblock \bibinfo{title}{{Handbook of global optimization. Volume 2}}.
\newblock \bibinfo{publisher}{Kluwer}.
\bibitem[{Press et~al.(1992)Press, Flannery, Teukolsky and
  Vetterling}]{press1992numerical}
\bibinfo{author}{Press, W.}, \bibinfo{author}{Flannery, B.},
  \bibinfo{author}{Teukolsky, S.}, \bibinfo{author}{Vetterling, W.},
  \bibinfo{year}{1992}.
\newblock \bibinfo{title}{{Numerical recipes in C: the art of scientific
  programming}}.
\newblock \bibinfo{journal}{Cambridge U. Press, Cambridge, England} .
\bibitem[{Saltelli(2002)}]{saltelli2002making}
\bibinfo{author}{Saltelli, A.}, \bibinfo{year}{2002}.
\newblock \bibinfo{title}{{Making best use of model evaluations to compute
  sensitivity indices}}.
\newblock \bibinfo{journal}{Computer Physics Communications}
  \bibinfo{volume}{145}, \bibinfo{pages}{280--297}.
\bibitem[{Saltelli et~al.()Saltelli, Chan and Scott}]{saltelli-sensitivity}
\bibinfo{author}{Saltelli, A.}, \bibinfo{author}{Chan, K.},
  \bibinfo{author}{Scott, E.}, .
\newblock \bibinfo{title}{{Sensitivity analysis. 2000}}.
\bibitem[{Sobol(2001)}]{sobol2001global}
\bibinfo{author}{Sobol, I.}, \bibinfo{year}{2001}.
\newblock \bibinfo{title}{{Global sensitivity indices for nonlinear
  mathematical models and their Monte Carlo estimates}}.
\newblock \bibinfo{journal}{Mathematics and Computers in Simulation}
  \bibinfo{volume}{55}, \bibinfo{pages}{271--280}.
\bibitem[{Storlie et~al.(2009)Storlie, Swiler, Helton and
  Sallaberry}]{storlie2009implementation}
\bibinfo{author}{Storlie, C.}, \bibinfo{author}{Swiler, L.},
  \bibinfo{author}{Helton, J.}, \bibinfo{author}{Sallaberry, C.},
  \bibinfo{year}{2009}.
\newblock \bibinfo{title}{{Implementation and evaluation of nonparametric
  regression procedures for sensitivity analysis of computationally demanding
  models}}.
\newblock \bibinfo{journal}{Reliability Engineering \& System Safety}
  \bibinfo{volume}{94}, \bibinfo{pages}{1735--1763}.
\bibitem[{Veroy and Patera(2005)}]{veroy2005certified}
\bibinfo{author}{Veroy, K.}, \bibinfo{author}{Patera, A.},
  \bibinfo{year}{2005}.
\newblock \bibinfo{title}{{Certified real-time solution of the parametrized
  steady incompressible Navier-Stokes equations: Rigorous reduced-basis a
  posteriori error bounds}}.
\newblock \bibinfo{journal}{International Journal for Numerical Methods in
  Fluids} \bibinfo{volume}{47}, \bibinfo{pages}{773--788}.

\end{thebibliography}

\end{document}